\documentclass[10pt,journal]{IEEEtran}
\usepackage{etex}
\usepackage[english]{babel}
\usepackage{amsmath}
\usepackage{gensymb}
\usepackage{psfrag}
\usepackage{graphicx}
\usepackage{url}
\usepackage{cite}
\usepackage{listings}
\usepackage{subfig}
\usepackage{units}
\usepackage[autolanguage]{numprint}
\usepackage{booktabs}
\usepackage[xindy,toc]{glossaries}
\usepackage[disable]{todonotes}
\usepackage{bytefield}
\usepackage{xspace}
\usepackage{microtype}
\usepackage{tikz}
\usepackage{tikz-timing}
\usepackage[final]{changes}
\usepackage[binary-units]{siunitx}

\nprounddigits{2}

\makeglossaries

\newacronym{adc}{ADC}{Analog to Digital Converter}
\newacronym{dac}{DAC}{Digital to Analog Converter}
\newacronym{ppu}{PPU}{Plasticity Processing Unit}
\newacronym{isa}{ISA}{Instruction Set Architecture}
\newacronym{fifo}{FIFO}{First In First Out}
\newacronym{fpga}{FPGA}{Field Programmable Gate Array}
\newacronym{fwhm}{FWHM}{full width at half maximum}
\newacronym{serdes}{SerDes}{Serializer/Deserializer}
\newacronym{simd}{SIMD}{Single Instruction Multiple Data}
\newacronym{sram}{SRAM}{Static Random Access Memory}
\newacronym{stdp}{STDP}{Spike-timing dependent plasticity}
\newacronym{io}{IO}{Input/Output}
\newacronym{dls}{HICANN-DLS}{HICANN-DLS}
\newacronym{lsb}{LSB}{least significant bit}
\newacronym{inl}{INL}{integral nonlinearity}
\newacronym{msb}{MSB}{most significant bit}
\newacronym{gpu}{GPU}{Graphics Processing Unit}
\newacronym{asic}{ASIC}{Application Specific Integrated Circuit}

\makeatletter
\renewcommand*{\@seccntformat}[1]{   \csname the#1\endcsname.\quad
}
\makeatother

\newcommand{\greatTitle}{Demonstrating Hybrid Learning in a Flexible Neuromorphic Hardware System}
\title{\greatTitle}

\author{Simon~Friedmann$^\dagger$
, Johannes~Schemmel$^\dagger$,~\IEEEmembership{Member,~IEEE}, Andreas~Gr\"ubl, Andreas~Hartel, Matthias~Hock and Karlheinz~Meier

\thanks{Authors marked with $^\dagger$
 contributed equally to this work.}
\thanks{Johannes Schemmel, Andreas Gr\"ubl, Andreas Hartel and Karlheinz Meier are with the Kirchhoff Institute for Physics, Heidelberg, Germany (e-mail: schemmel$@$kip.uni-heidelberg.de).}
\thanks{Simon Friedmann and Matthias Hock were with the Kirchhoff Institute for Physics, Heidelberg, Germany.}
}
\begin{document}
\maketitle

\begin{abstract}
         We present results from a new approach to learning and plasticity in neuromorphic hardware systems:
    to enable flexibility in implementable learning mechanisms while keeping high efficiency associated with neuromorphic implementations, we combine a general-purpose processor with full-custom analog elements.
    This processor is operating in parallel with a fully parallel neuromorphic system consisting of an array of synapses connected to analog, continuous time neuron circuits. Novel analog correlation sensor circuits process spike events for each synapse in parallel and in real-time.   
    The processor uses this pre-processing to compute new weights possibly using additional information following its program.
    Therefore, \added{to a certain extent,} learning rules can be defined in software giving a large degree of flexibility.
    Synapses realize correlation detection geared towards Spike-Timing Dependent Plasticity (STDP) as central computational primitive in the analog domain.
    Operating at a speed-up factor of 1000 compared to biological time-scale, we measure time-constants from tens to hundreds of micro-seconds.
    We analyze variability across multiple chips and demonstrate learning using a multiplicative STDP rule.
    We conclude that the presented approach will enable flexible and efficient learning as a platform for neuroscientific research and technological applications.
    \footnote{DOI$:$ 10.1109$/$TBCAS.2016.2579164}
    \footnote{IEEE explore open access: \url{http://ieeexplore.ieee.org/document/7563782}}
    \footnote{\copyright 2016 IEEE. Translations and content mining are permitted for academic research only. Personal use is also permitted, but republication/redistribution requires IEEE permission. See \url{http://www.ieee.org/publications_standards/publications/rights/index.html} for more information.}

\end{abstract}

\begin{IEEEkeywords}
    digital signal processing, learning, synapse circuit, neuromorphic hardware, spike-time dependent plasticity
\end{IEEEkeywords}

\IEEEpeerreviewmaketitle

\newcommand{\usableVstoreMin}{0.310000\xspace}
\newcommand{\usableVstoreMinP}{\numprint{0.310000}\xspace}
\newcommand{\usableVstoreMinAmplCausalErr}{0.052350\xspace}
\newcommand{\usableVstoreMinAmplCausalErrP}{\numprint{0.052350}\xspace}
\newcommand{\usableVstoreMinAmplCausalMean}{0.909326\xspace}
\newcommand{\usableVstoreMinAmplCausalMeanP}{\numprint{0.909326}\xspace}
\newcommand{\variationOffset}{1.000000\xspace}
\newcommand{\variationOffsetP}{\numprint{1.000000}\xspace}
\newcommand{\usableVstoreMax}{0.395000\xspace}
\newcommand{\usableVstoreMaxP}{\numprint{0.395000}\xspace}
\newcommand{\ppuTauDiffErr}{2.588379\xspace}
\newcommand{\ppuTauDiffErrP}{\numprint{2.588379}\xspace}
\newcommand{\usableVrampMaxFwhmCausalMean}{11.363636\xspace}
\newcommand{\usableVrampMaxFwhmCausalMeanP}{\numprint{11.363636}\xspace}
\newcommand{\ppuTauDiffMean}{2.102148\xspace}
\newcommand{\ppuTauDiffMeanP}{\numprint{2.102148}\xspace}
\newcommand{\usableVstoreMaxAmplCausalErr}{0.055658\xspace}
\newcommand{\usableVstoreMaxAmplCausalErrP}{\numprint{0.055658}\xspace}
\newcommand{\usableVstoreMinAmplMean}{0.910955\xspace}
\newcommand{\usableVstoreMinAmplMeanP}{\numprint{0.910955}\xspace}
\newcommand{\usableVstoreMaxAmplMean}{0.146412\xspace}
\newcommand{\usableVstoreMaxAmplMeanP}{\numprint{0.146412}\xspace}
\newcommand{\usableVrampMax}{0.270000\xspace}
\newcommand{\usableVrampMaxP}{\numprint{0.270000}\xspace}
\newcommand{\usableVrampMaxFwhmErr}{3.565558\xspace}
\newcommand{\usableVrampMaxFwhmErrP}{\numprint{3.565558}\xspace}
\newcommand{\usableVrampMinFwhmMean}{228.731762\xspace}
\newcommand{\usableVrampMinFwhmMeanP}{\numprint{228.731762}\xspace}
\newcommand{\usableVrampMaxFwhmMean}{12.310606\xspace}
\newcommand{\usableVrampMaxFwhmMeanP}{\numprint{12.310606}\xspace}
\newcommand{\usableVrampMinFwhmCausalErr}{108.428545\xspace}
\newcommand{\usableVrampMinFwhmCausalErrP}{\numprint{108.428545}\xspace}
\newcommand{\usableVrampMinFwhmErr}{85.301553\xspace}
\newcommand{\usableVrampMinFwhmErrP}{\numprint{85.301553}\xspace}
\newcommand{\usableVstoreMinAmplErr}{0.142902\xspace}
\newcommand{\usableVstoreMinAmplErrP}{\numprint{0.142902}\xspace}
\newcommand{\usableVstoreMaxAmplCausalMean}{0.198317\xspace}
\newcommand{\usableVstoreMaxAmplCausalMeanP}{\numprint{0.198317}\xspace}
\newcommand{\variationNumSynapsesInC}{32\xspace}
\newcommand{\variationNumSynapsesInCP}{32\xspace}
\newcommand{\variationNumSynapsesInB}{800\xspace}
\newcommand{\variationNumSynapsesInBP}{800\xspace}
\newcommand{\variationNumSynapsesInA}{672\xspace}
\newcommand{\variationNumSynapsesInAP}{672\xspace}
\newcommand{\variationNumSynapsesInD}{32\xspace}
\newcommand{\variationNumSynapsesInDP}{32\xspace}
\newcommand{\usableVrampMinFwhmCausalMean}{336.595118\xspace}
\newcommand{\usableVrampMinFwhmCausalMeanP}{\numprint{336.595118}\xspace}
\newcommand{\usableVrampMin}{0.160000\xspace}
\newcommand{\usableVrampMinP}{\numprint{0.160000}\xspace}
\newcommand{\usableVstoreMaxAmplErr}{0.037800\xspace}
\newcommand{\usableVstoreMaxAmplErrP}{\numprint{0.037800}\xspace}
\newcommand{\usableVrampMaxFwhmCausalErr}{3.083230\xspace}
\newcommand{\usableVrampMaxFwhmCausalErrP}{\numprint{3.083230}\xspace}

\newcommand{\dacSlope}{11.516865\xspace}
\newcommand{\dacSlopeP}{\numprint{11.516865}\xspace}
\newcommand{\dacOffsetErr}{0.000158\xspace}
\newcommand{\dacOffsetErrP}{\numprint{0.000158}\xspace}
\newcommand{\dacOffset}{22.786151\xspace}
\newcommand{\dacOffsetP}{\numprint{22.786151}\xspace}
\newcommand{\dacSlopeErr}{0.000000\xspace}
\newcommand{\dacSlopeErrP}{\numprint{0.000000}\xspace}
\newcommand{\dacInl}{4.834144\xspace}
\newcommand{\dacInlP}{\numprint{4.834144}\xspace}
\newcommand{\dacMeanInl}{1.056345\xspace}
\newcommand{\dacMeanInlP}{\numprint{1.056345}\xspace}

\section{Introduction}
\label{sec:intro}

In the modern landscape of information technology machine learning is gaining more and more in importance.
Major companies use artificial intelligence for their products \cite{lecun2015deep}.
This development is driven by advancements in methods such as deep learning \cite{schmidhuber1992learning,hinton2007learning} that were originally inspired by concepts from neuroscience.
Together with the availability of substantial computational performance, these methods enable complex machine learning applications, such as image \cite{krizhevsky2012imagenet} or speech recognition \cite{hinton2012deep}.
Specialized hardware can lower the cost of these methods in terms of energy, time, and therefore money \cite{ovtcharov2015accelerating}, enabling either a scaling to larger problem sizes or the use in new devices outside of data centers.

\replaced{On the other hand, using simulations of neural networks as a major tool for research in neuroscience depends on efficient simulators for large-scale networks.}{On the other hand, neuroscience using simulations of neural networks as a major tool for research depends on efficient simulators for large-scale networks.}
This opens the opportunity to build specialized hardware systems that serve as efficient platforms for research as well as technology.
Multiple systems with this goal have been proposed, e.g.\ \cite{schemmel2010iscas,furber2012,merolla2014million,qiao2015reconfigurable}.

While the problem can be approached in different ways, the concept of analog neuromorphic hardware  \cite{mead90neuromorphic,douglas95neuromorphic} promises especially area and energy efficient solutions as demonstrated by e.g.\ \cite{indiveri2003low,schemmel_ijcnn06,wijekoon2008}.
These systems use the concept of a physical model to emulate neural networks: the temporal development of the membrane voltages of the neurons is emulated by custom analog circuits, representing the neuron and synapses of the emulated network.
However, neurons and synapses built this way are limited to at best a family of models that are compatible with their physical realization.
On the other end of the spectrum, software allows the simulation of arbitrary models by solving numerical equations.

\replaced{Especially, there exists a large set of different models for learning and plasticity, so that a flexible hardware implementation is desirable.}{Especially for learning and plasticity exist a large set of different models, so that a flexible hardware implementation is desirable.}
This is true for technical applications where one network is often trained with different methods for pre-training and fine-tuning \cite{hinton2007learning}, as well as biology where different plasticity rules are found depending on cell type and brain region \cite{abbott2000synaptic,caporale08_stdp}.
But besides flexibility, efficiency is a key concern in both domains.
Large-scale simulations have been demonstrated in the past \cite{Ananthanarayanan2009,helias2012supercomputers,markram2015reconstruction}, but, especially with plasticity, simulation time quickly becomes a limiting factor even on medium-sized networks \cite{zenke2014auryn} .
Similarly, in the technical domain, significant effort is put into accelerating learning including the use of \glspl{gpu} and \glspl{fpga} \cite{chilimbi2014project,ovtcharov2015accelerating}.

For this study, we follow a novel hybrid approach to learning as a trade-off between efficiency and flexibility:
we use full-custom analog circuits for real-time and parallel processing of spikes in the emulated synapses.
These circuits serve as sensors for an embedded general-purpose processor that implements the learning rule in software.
This way, we offer a solution that allows biologically realistic plasticity while emulating networks a thousand times faster than in biology.
Using physical models for core components, this speed-up is not affected by network size or activity.
In this study we present results from a scaled-down prototype that demonstrates for the first time plasticity in such a hybrid system using analog components together with an embedded \gls{ppu}.

The study starts with a description of analog circuits and the architecture of the \gls{ppu} in Section~\ref{sec:circuits}.
After that, we introduce the theoretical background and methods in Section~\ref{sec:theory}.
Then, results are presented for simulations in Section~\ref{sec:sim} and for experiments in Section~\ref{sec:experiments}.
Finally, Section~\ref{sec:discussion} discusses results, followed by conclusion and outlook in Sections~\ref{sec:conclusion} and~\ref{sec:outlook}.

\section{Description of Circuits}
\label{sec:circuits}
\begin{figure}
    \centering
    \includegraphics[width=85mm]{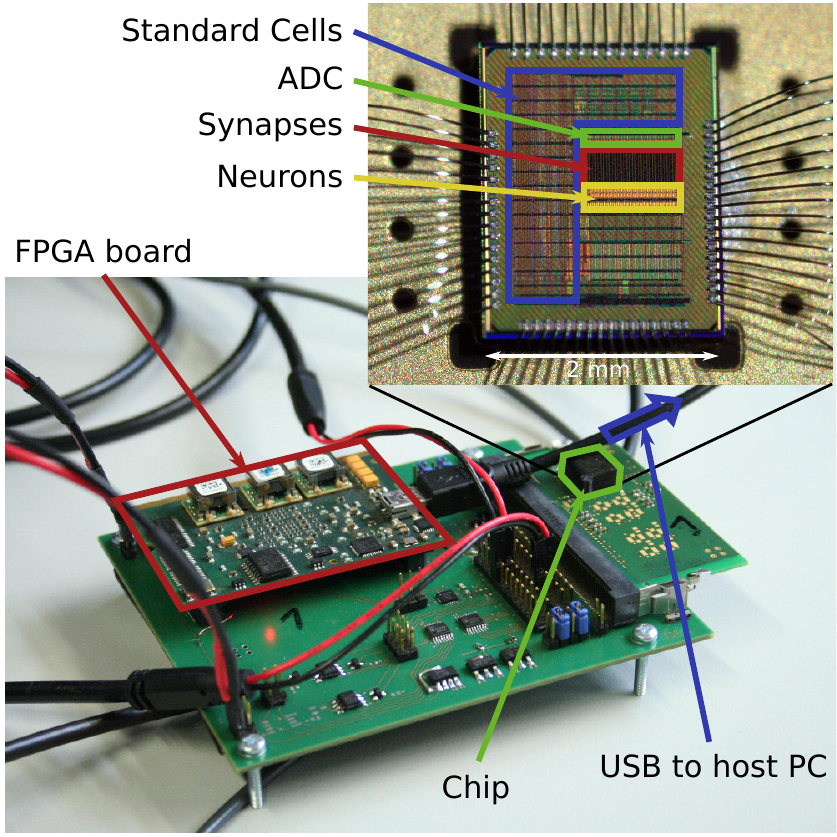}
    \caption{        Photograph of die and test system. The active die area is \unit{1.7$\times$2.2}{mm$^2$}.
        The host computer communicates via USB with an \gls{fpga} board.
        The \gls{fpga} controls \glspl{dac} on the board for bias generation and communicates with the chip through a \gls{serdes} interface.
    }
    \label{fig:board}
\end{figure}
\begin{figure}
    \center{\includegraphics[width=0.98\linewidth,page=5,viewport=0 0 28cm 19cm, clip]{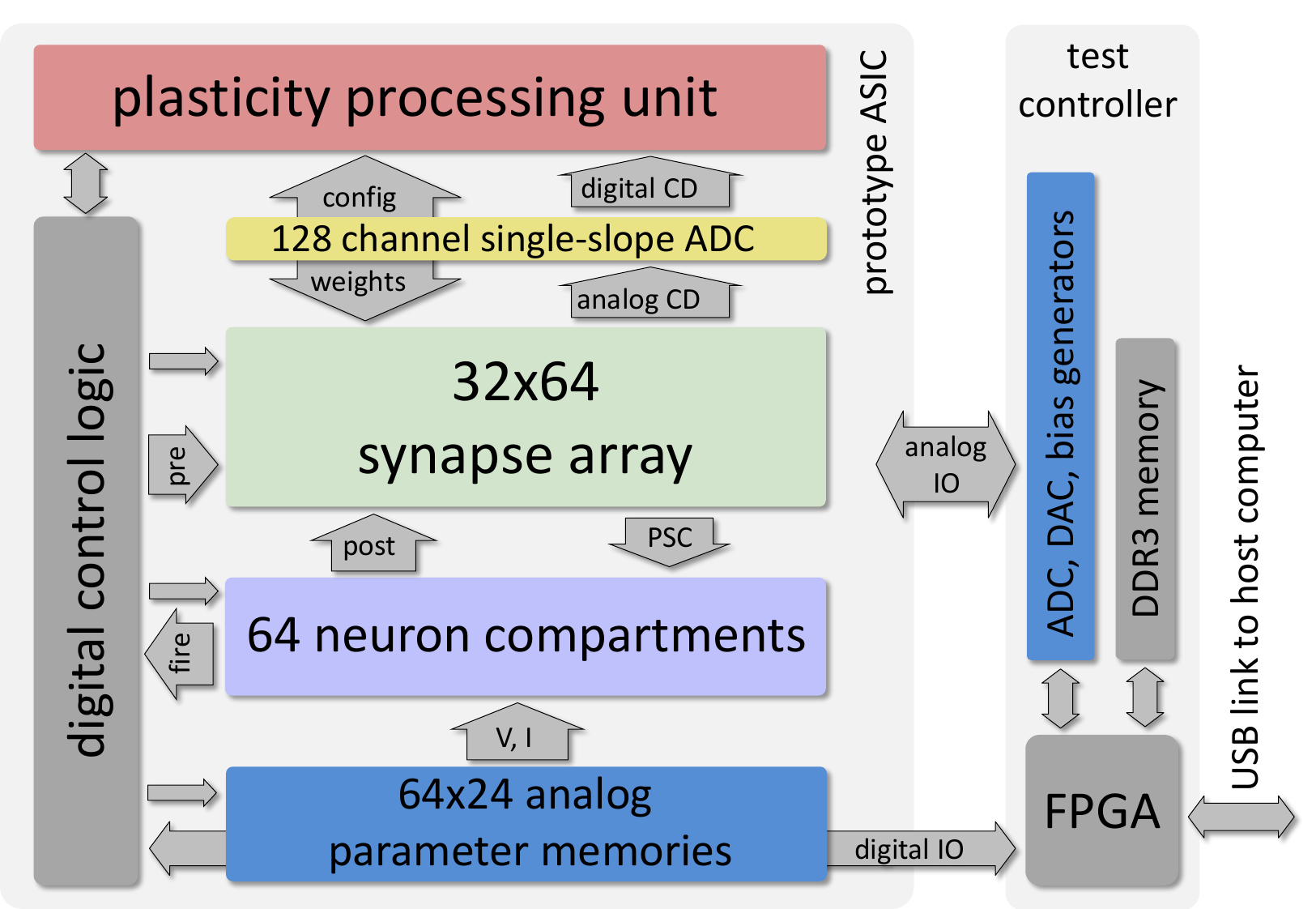}  
    \caption{Block diagram of the presented system. The prototype ASIC is shown to the left. A photograph of the system can be seen in Fig.~\ref{fig:board}. 'CD' stands for 'correlation data' and 'PSC' for 'post-synaptic current'.}
    \label{fig:overview} }
\end{figure}

The circuits presented in this paper are part of a prototype ASIC for the next generation of a large Neuromorphic Hardware system \cite{schemmel2010iscas}. All results have been measured using the setup shown in Fig.~\ref{fig:board}. The individual components of the chip and their functional relations are depicted in Fig.~\ref{fig:overview}. The central elements are an array of 2048 synapses and 64 neuron-compartment circuits, which implement the analog, continuous-time emulation of their biological counterparts. Similar to the predecessor system described in \cite{schemmel2010iscas} the presented chip operates faster than \replaced{wall-clock time}{wall-time}. To simplify the calibration of the analog elements to the model equations, the acceleration factor is fixed at $10^3$. Therefore, one second in the model time scale is emulated in one millisecond by \replaced{the presented system}{this work}.

The focus of this paper is the plasticity sub-system, which observes the activity of the emulated neural network and modifies its parameters in reaction to these observations depending on the configured plasticity rule. The neuron circuits are not covered in this publication.

The plasticity sub-system is a mixed-signal, highly-parallel control loop simultaneously monitoring the temporal correlation between all pre- and post-synaptic firing times.
The plasticity rule itself is implemented as software running on an embedded micro-processor, the \gls{ppu}.
It evaluates the signals from the analog correlation sensors located within the synapses and computes weight updates.
Besides the synapse, it can observe firing rates of neurons and modify parameters of the emulated neurons as well as the topology of the network. Connection to the outside world allows the integration of third factors, for example a reward signal \cite{Friedmann2013}.

The parallel analog implementation of the correlation sensors in every synapse allows the plasticity sub-system to handle the high rate of simultaneous events\footnote{Due to the acceleration factor of $10^3$ every component has to handle a thousandfold higher data rate as a comparable \replaced{unaccelerated system operating at biological time scale}{real-time system}.} The circuit maintains a local eligibility trace that depends on the relative timing of pre- and post-synaptic firing.

A 128 channel single-slope \gls{adc} \footnote{The initial design of the \gls{adc} was done by Sabanci University, Turkey.} digitizes the stored trace information for the \gls{ppu}.

\subsection{Synapse}
\label{sec:circ:synapse}

\subsubsection{Basic Operation Principles}
In Fig.~\ref{fig:overview} the synapses are arranged in a two-dimensional array between the \gls{ppu} and the neuron compartment circuits. Pre-synaptic input enters the synapse array at the left edge. For each row, a set of signal buffers transmit the pre-synaptic pulses to all synapses in the row. The post-synaptic side of the synapses, i.e. the equivalent of the dendritic membrane of the target neuron, is formed by wires running vertically through each column of synapses. 

At each intersection between pre- and post-synaptic wires, a synapse is located. To avoid that all neuron compartments share the same set of pre-synaptic inputs, each pre-synaptic input line transmits - in a time-multiplexed fashion - the pre-synaptic signals of up to 64 different pre-synaptic neurons. Each synapse stores a pre-synaptic address that determines the pre-synaptic neuron it responds to.

\begin{figure}
	\center{\includegraphics[width=0.98\linewidth,page=1,viewport=0 0 27cm 12cm, clip]{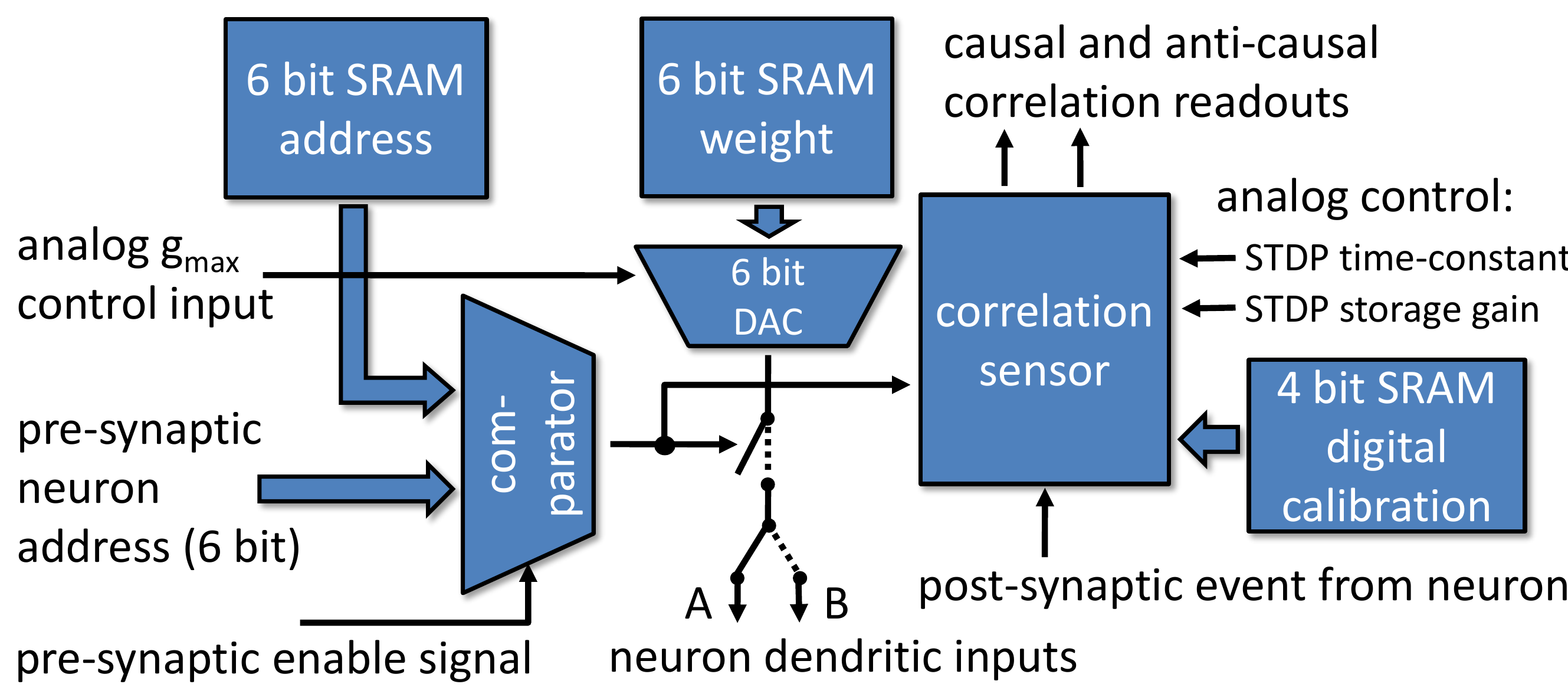}
	\caption{Block diagram of the synapse circuit.}
	\label{fig:synapse_block} }
\end{figure}
Fig.~\ref{fig:synapse_block} shows a block diagram of the synapse circuit. The main functional blocks are the address comparator, the \gls{dac} and the correlation sensor. Each of these circuits has its associated memory block.

The address comparator receives a \unit[6]{bit} address and a pre-synaptic enable signal \added{from the periphery of the synapse array as well as a locally stored \unit[6]{bit} neuron number.}
If the address matches the programmed neuron number, the comparator circuit generates a pre-synaptic enable signal local to the synapse ({\em pre}), which is subsequently used in the \gls{dac} and correlation sensor circuits. 

Each time the \gls{dac} circuit receives a {\em pre} signal, it generates a current pulse. The height of this pulse is proportional to the stored weight, while the pulse width is typically \unit{4}{ns}. This matches the maximum pre-synaptic input rate of the whole synapse row which is limited to \unit{125}{MHz}. The remaining \unit{4}{ns} are necessary to change the pre-synaptic address. The current pulse can be shortened below the \unit{4}{ns} maximum pulse length to emulate short-term synaptic plasticity \cite{schemmel_iscas07}.

Each neuron compartment has two inputs, labeled A and B in Fig.~\ref{fig:synapse_block}. Usually, the neuron compartment uses A as excitatory and B as inhibitory input. Each row of synapses is statically switched to either input A or B, meaning that all pre-synaptic neurons connected to this row act either as excitatory or inhibitory inputs to their target neurons. Due to the address width of \unit[4]{bit} the maximum number of different pre-synaptic neurons is 64.

The remaining block shown in Fig.~\ref{fig:synapse_block} is the correlation sensor, which has a \unit[4]{bit} static memory associated with it. Its task is the measurement of the time difference between pre- and post-synaptic spikes. To determine the time of the pre-synaptic spike it is connected to the {\em pre} signal. The post-synaptic spike-time is determined by a dedicated signaling line running from each neuron compartment vertically through the synapse array to connect to all synapses projecting to inputs A or B of the compartment. This signal, which is called {\em post} subsequently, has a similar pulse length as the {\em pre} signal.

The correlation sensor measures the causal ({\em pre} before {\em post}) and anti-causal ({\em post} before {\em pre}) time differences and stores them as exponentially weighted sums within the synapse circuit.
In comparison to earlier implementations \cite{schemmel_ijcnn06} by the authors the circuit has been improved in two main aspects: first, only one instance of the time measurement circuit is now re-used for causal as well as anti-causal time difference measurements, resulting in strongly reduced mismatch between the causal and anti-causal branches of the activation function. Second, the time-constant of the exponential is now truly adjustable over more than two orders of magnitude to fit most biological models of spike-time dependent plasticity \cite{dan06stdp}.

\begin{table}
    \centering
    \caption{Key parameters of the synapse circuit}
    \label{tab:synparam}
    \begin{tabular}{cc}
        \toprule
        \textbf{Parameter} & \textbf{Value}  \\
        \midrule
        $V_\textrm{dd}$ thin oxide & \unit{1.2}{V} \\     
        $V_\textrm{dd}$ thick oxide & \unit{2.5}{V} \\
        area & \unit{94}{$\mu m^2$} \\
        total \# of MOSFET & 205 \\
        $C_\textrm{causal, anti-causal}$ & \unit{6}{fF}, MOSCAP\\
        $C_\textrm{transfer}$ & \unit{3-9}{fF}, MOSCAP, adjustable\\
        $C_\textrm{storage}$ & \unit{37}{fF}, MIMCAP\\
         \bottomrule
    \end{tabular}
\end{table}

Due to the implementation in a much smaller process feature size, \unit{65}{nm} instead of \unit{180}{nm}, four static memory bits could be allocated for additional calibration of transistor variations within each synapse. \added{Table~\ref{tab:synparam} summarizes key parameters of the synapse implementation.}
\begin{figure}
	\center{\includegraphics[width=0.98\linewidth,page=4,viewport=0 0 27cm 10.5cm, clip]{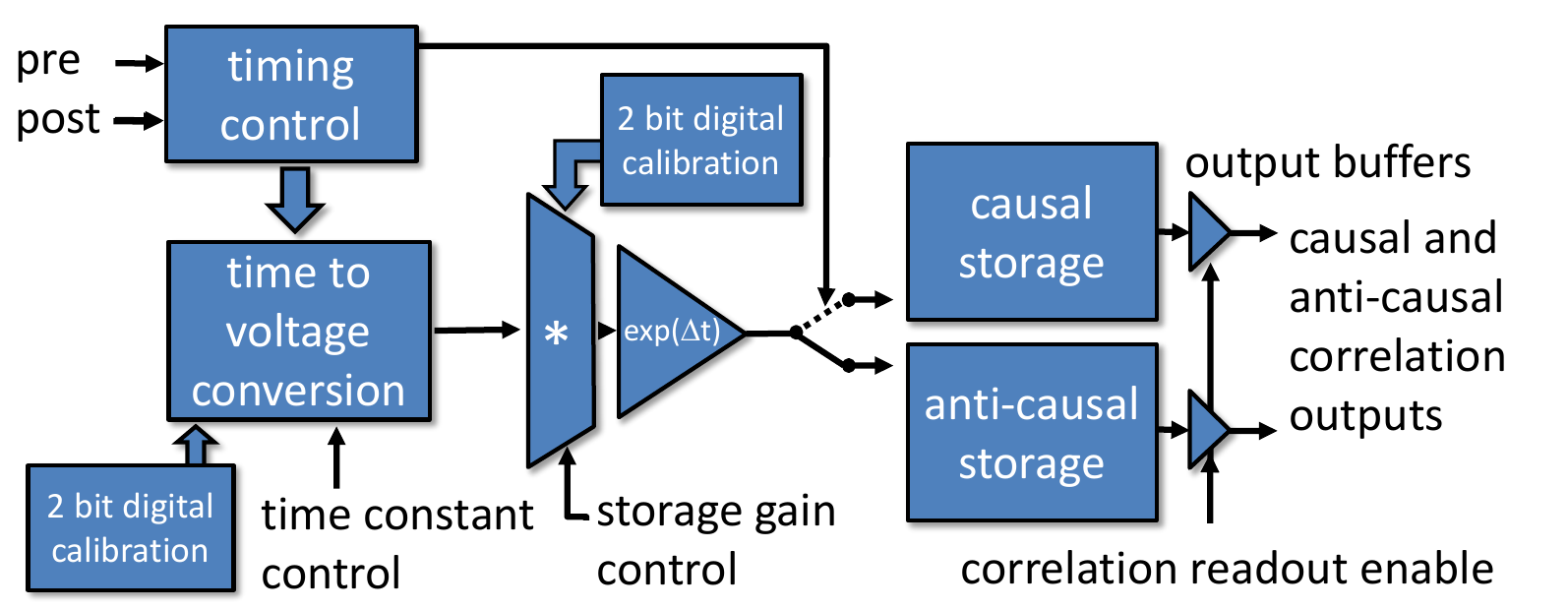}
	\caption{Block diagram of the correlation sensor circuit.}
	\label{fig:corrsense_block} }
\end{figure}

\subsubsection{Correlation Sensor Circuit}
The structure of the correlation sensor is shown in Fig.~\ref{fig:corrsense_block}. The input stage receives {\em pre} and {\em post} signals and uses them to generate the internal timing. A time to voltage conversion circuit generates a voltage representing the elapsed time between the most recent {\em pre} and {\em post} events. This voltage is scaled by the storage gain parameter and the result is used as argument to an exponential function. This exponentially weighted time difference is added to one of two storage circuits. The selection of the storage circuit depends whether the last input event seen has been a {\em pre} or {\em post} signal. {\em Pre} before {\em post} is stored in the causal storage, {\em post} before {\em pre} in the anti-causal one. 

To counteract the effects of fixed-pattern noise created by transistor variations\added{,} the time to voltage as well as the storage gain stages have two digital calibration inputs each. The four calibration bits are stored locally in each synapse. The time constant of the time to voltage conversion can be set for one row of synapses by a control voltage. The same applies to the storage gain stage, where the storage gain control signal adjusts one row of synapses.
In the prototype chip the gain and time constant input signals of each row are shorted and connected to two external input pins.

The values stored in the causal and the anti-causal storage cells can be read out simultaneously for all synapses in a row. A parallel single-slope \gls{adc} at the top of the synapse array converts the analog values read out from the storage cells into digital words for the \gls{ppu} (see Fig.~\ref{fig:overview}).

\begin{figure*}
	\center{\includegraphics[width=0.9\linewidth,page=2,viewport=0 0 29cm 19cm, clip]{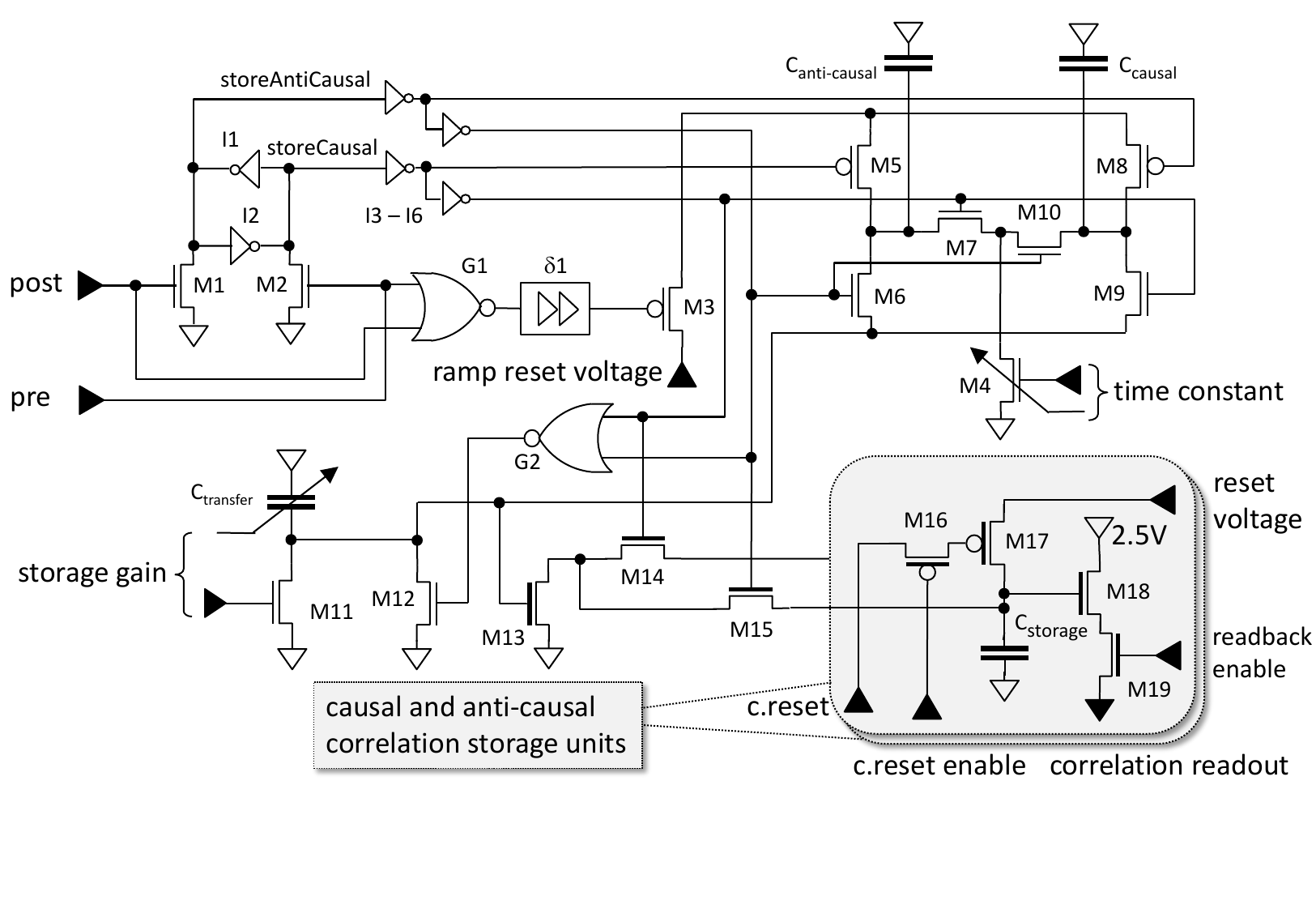}
	\caption{Simplified circuit diagram of the correlation sensor. All supply connections are \unit{1.2}{V} if not stated otherwise. A thick gate symbol depicts a thick gate-oxide transistor capable of \unit{2.5}{V} operation. The assignment of the components to the functional blocks depicted in Fig.~\ref{fig:corrsense_block} is as follows: timing control - M1-2, I1-6, G1-2, $\delta 1$; time to voltage conversion - M3-10, $C_\textrm{causal}$, $C_\textrm{anti-causal}$; exponential - M13; storage gain - M11-12, $C_\textrm{transfer}$; storage - M14-17; storage output buffer - M18-19, $C_\textrm{storage}$.\label{fig:corrsense_circuit} }
	}
\end{figure*}
Fig.~\ref{fig:corrsense_circuit} depicts the correlation sensor circuit. To enhance the readability of the circuit diagram, the individual blocks of Fig.~\ref{fig:corrsense_block} are not marked. See the caption for assignments of the components to the different functional blocks.

\begin{figure}
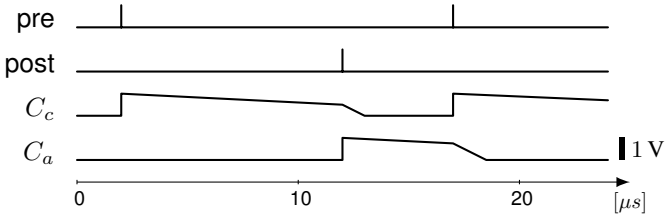

	\begin{center}
		\begin{tikztimingtable} 
			pre & 2{L} G 15{L} G 7{L}\\
			post & 12{L} G 12{L} \\
			$C_c$ & 2{L} !{-- +(0,1) -- +(10,.5) -- ++(11,0)} 4{L} !{-- +(0,1) -- ++(7,.7)}\\ 
			$C_a$ & 12{L} !{-- +(0,1) -- +(5,.75) -- ++(6.5,0)} 5{L} .5L\\
		\begin{extracode}
									\path node[label={[label distance=-1mm]right:\small\SI{1}{V}}]
			(scale) at (+24.5,-5.5) {};
									\fill[color=black] ($ (scale) + (0,0.5)$) rectangle +(0.25,-1);
			\begin{background}
		  				  				  				  		\draw [->,>=latex] (0,-\nrows-3) -- (\twidth+1,-\nrows-3)
		  		node [draw=none,below,align=left]{\scalebox{.8}{$\left[\mu s\right]$}};
		  		\foreach \n in {0,10,...,\twidth}
		    	\draw (\n,-\nrows-3+.1) -- +(0,-.2)
		        node [below,inner sep=2pt] {\scalebox{.75}{		        \n}};

		 	\end{background}
		 			\end{extracode}
		\end{tikztimingtable} 
		\caption{Exemplary timing of the correlation sensor circuit. The timescale is of the order of the correlation sensor time constant $\tau_{c}$.\label{fig:timing_corrsensor_prepost}}
	
	\end{center}
\end{figure}
As stated above, the correlation sensor monitors the temporal correlation between {\em pre} and {\em post} synaptic firing events. 
This is accomplished by charging the capacitors $C_\textrm{causal}$ and $C_\textrm{anti-causal}$ with a constant current. The selection of the capacitor depends on the temporal order of the pre and post signals. As can be seen in Fig.~\ref{fig:timing_corrsensor_prepost}, the arrival of a {\em pre} pulse starts the charging of $C_\textrm{causal}$ after discharging it quickly to its initial value, while $C_\textrm{anti-causal}$ starts charging after a {\em post} pulse. Two or more {\em pre} or {\em post} pulses in succession would only restart the discharge/charge process without changing the capacitor. Therefore, the correlation sensor only supports plasticity rules based on nearest neighbor schemes \cite{morrison08_stdp}.

\begin{figure}
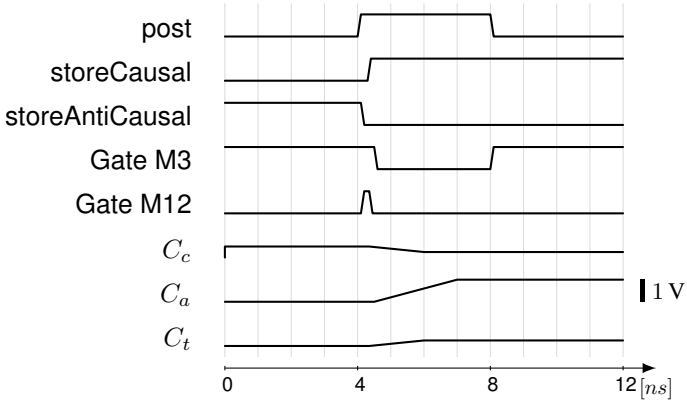

	\begin{center}
		\begin{tikztimingtable} [xscale=1.5]
		 	post  &  4{L} HHHH 4{L}   \\
 			storeCausal & 4{L} 0.3L 0.7H 7{H}\\
 			storeAntiCausal & 4{H} .1H 0.9L 7{L}\\
 			Gate M3  &  4{H} hl 3{L} 4{H}  \\	
 			Gate M12 &  4{L} 0.1L 0.25H 0.65L 7{L}\\
 			$C_c$ & !{-- +(0,0.5) -- +(4.35,0.5) -- +(6,0.25) -- ++(12,0.25)}  \\
 			$C_a$ & !{-- +(0,0) -- +(4.5,0) -- +(7,1) -- ++(12,1)}\\ 			
 			$C_t$ & !{-- +(0,0) -- +(4.35,0) -- +(6,0.25) -- ++(12,0.25)} \\
 			\begin{extracode}
												\path node[label={[label distance=-1mm]right:\small\SI{1}{V}}]
				(scale) at (+12.5,-11.5) {};
												\fill[color=black] ($ (scale) + (0,0.5)$) rectangle +(0.15,-1);

				\begin{background}
			  		\vertlines[help lines,opacity=0.3]{}
			  					  		
			  		\draw [->,>=latex] (0,-\nrows-7) -- (\twidth+1,-\nrows-7)
			  		node [draw=none,below,align=left]{\scalebox{.8}{$\left[ ns \right]$}};
			  		\foreach \n in {0,4,...,\twidth}
			    	\draw (\n,-\nrows-7+.1) -- +(0,-.2)
			        node [below,inner sep=2pt] {\scalebox{.75}{			        \n}};
			 	\end{background}
			 				\end{extracode}
		\end{tikztimingtable} 

	\caption{Timing of the correlation sensor circuit: zoom-in on the time axis around the post event shown in Fig.~\ref{fig:timing_corrsensor_prepost}. The following abbreviations are used for the capacitor labels: $C_\textrm{c(ausal)} $, $C_\textrm{a(nti-causal)}$ and $C_\textrm{t(ransfer)}$.\label{fig:timing_corrsensor_zoompost}}
	
	\end{center}
\end{figure}

To determine the temporal order, the input stage of the correlation sensor utilizes a D-latch formed by I1 and I2. Each time a {\em post} follows a {\em pre} or vice-versa, the D-latch gets toggled by M1 or M2, respectively. To orchestrate the precise discharging and switching of capacitors within the limited area of the synapse, the circuit makes use of the delays of the individual components. In Fig.~\ref{fig:timing_corrsensor_zoompost} a subset of the relevant signals is shown. 
The inverters I1 and I2, which form the D-latch, have a very low drive strength. This leads to a significant delay between the internal node being discharged by an external {\em pre} or {\em post} pulse, and the respective inverted internal node ({\em storeAntiCausal} in case of a {\em pre} pulse or {\em storeCausal} after a {\em post} signal).

This time difference is used by G2 to produce a short pulse at the gate of M12 to precharge $C_\textrm{transfer}$ (see below).
\begin{figure}
	\center{\includegraphics[width=0.9\linewidth]{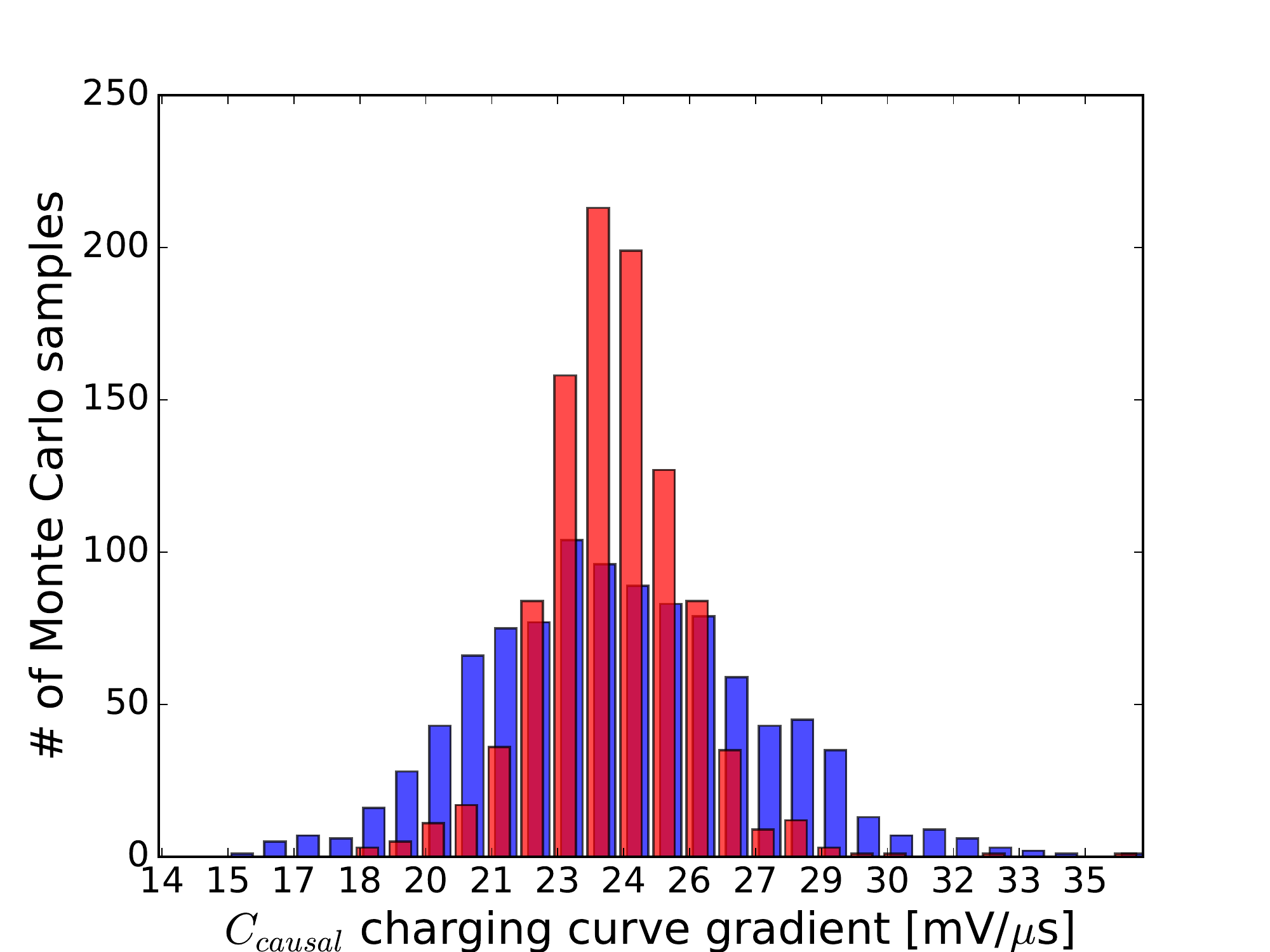} 	\caption{\added{Results of a Monte-Carlo simulation showing the effect of the two-bit digital time-constant calibration built into each synapse. The two histograms show the distribution of the gradient of the charging curve of $C_\textrm{causal}$ (dashed trace in Fig.~\ref{fig:sim_ctransdischarge}) over 1000 MC-runs each. The uncalibrated case is shown in blue, the calibrated in red.}}
	\label{fig:mchistvcausal} }
\end{figure}
The current charging the capacitors $C_\textrm{causal}$ and $C_\textrm{anti-causal}$, and therefore controlling the time constant of the correlation sensor, is generated by an adjustable current sink M4. The gate voltage of M4 is shared by all synapses of a row. To reduce the fixed pattern noise within a synapse row, the length of M4 can be digitally controlled in four steps by approximately \unit{20}{\%}.
This allows to reduce the fixed pattern noise by selecting for each synapse the value which minimizes synapse to synapse variation within the row.
\added{Fig.~\ref{fig:mchistvcausal} shows the results of a Monte-Carlo simulation demonstrating the effectiveness of this approach.}

In the full-size neural network chip each row will have an individual bias generation for M4, which allows different time constants in different rows, as well as the calibration of the row mean of the time-constant. The presented prototype chip directly connects all time-constant inputs to an external input pin which is driven by the test controller (see Fig.~\ref{fig:overview}).

The state of the D-latch determines whether $C_\textrm{causal}$ or $C_\textrm{anti-causal}$ is charged by M4 through the inverter chains formed by I3 to I6 and M7 as well as M10. 

The subsequent discussion is based on the temporal relations depicted in Fig.~\ref{fig:timing_corrsensor_zoompost}. As can be seen in Fig.~\ref{fig:timing_corrsensor_prepost}, the charging process of $C_\textrm{causal}$ or $C_\textrm{anti-causal}$ starts after it has been discharged to the ramp reset voltage by the {\em pre} or {\em post} event. In the case shown in Fig.~\ref{fig:timing_corrsensor_zoompost}, $C_\textrm{anti-causal}$ is discharged. The initial discharge is initated in two steps: first, after \deleted{the} arrival of a {\em post} pulse, $C_\textrm{anti-causal}$ is connected to M3 by enabling M5. The enabling of M3 is delayed to make sure the other capacitor, $C_\textrm{causal}$ is disconnected from M3 by M8. This is essential since at this moment $C_\textrm{causal}$ holds the last causal time-difference measurement result which should not be altered by the discharge of $C_\textrm{anti-causal}$.
\added{
		At the beginning of the {\em post} pulse the voltage on $C_\textrm{causal}$ is as follows:}
	\begin{align}
		V_{C_\textrm{causal}}^\textrm{begin post} &= V_\textrm{ramp reset} - \frac{  I_\textrm{M4} \cdot (t_\textrm{post}-t_\textrm{pre}) }{C_\textrm{causal}} \label{eq:vcausalanticausal}
	\end{align}
\added{After a {\em pre} pulse a similar equation holds for the voltage on $C_\textrm{anti-causal}$:}
	\begin{align}
		V_{C_\textrm{anti-causal}}^\textrm{begin pre} &= V_\textrm{ramp reset} - \frac{  I_\textrm{M4} \cdot (t_\textrm{pre}-t_\textrm{post}) }{C_\textrm{anti-causal}}
	\end{align}

The intial discharge process finishes within the time-interval set by the length of the {\em post} pulse. After {\em post} becomes inactive, M3 is deactivated and the charging of $C_\textrm{anti-causal}$ by the current flowing through M7 and M4 starts. Simultaneously, the transfer of the causal result from $C_\textrm{causal}$ to the storage capacitor $C_\textrm{storage causal}$ is initiated. In Fig.~\ref{fig:corrsense_circuit} only one of the two identical storage circuits is drawn. Depending on the state of the {\em storeCausal} and {\em storeAntiCausal} signals, M14 or M15 connect one of the storage circuits to M13. The timing of these signals assures that M14 and M15 are never activated simultaneously. 

The charge transfer starts by enabling M9, thereby connecting $C_\textrm{causal}$ to $C_\textrm{transfer}$. To avoid any crosstalk from the previous transfer process, M12 is always activated prior to M9 and charges $C_\textrm{transfer}$ to $V_\textrm{dd}$. After M9 is enabled, charge charing between 
$C_\textrm{causal}$ and $C_\textrm{transfer}$ starts. The charging process will be completed before {\em post} becomes inactive, but $C_\textrm{causal}$ and $C_\textrm{transfer}$ will stay connected until the end of the storage cycle.

\added{	 After the {\em post} pulse, before the charging of $C_\textrm{transfer}$ and $C_\textrm{causal}$ starts, the voltage on $C_\textrm{transfer}$ can be calculated as follows:}
	\begin{align}
		V_{C_\textrm{transfer}}^\textrm{end post} &= \frac{V_{C_\textrm{causal}}^\textrm{begin post} C_\textrm{causal} + V_{C_\textrm{transfer}}^\textrm{begin post} C_\textrm{transfer}}{C_\textrm{transfer}+C_\textrm{casual}}\label{eq:vtransferafterpost}
	\end{align}
\added{Since $C_\textrm{transfer}$ has been charged by M12 at the very beginning of the {\em post} pulse, $V_{C_\textrm{transfer}}^\textrm{begin post}$ is zero and Eq.~\ref{eq:vtransferafterpost} simplifies to: }
	\begin{align}
		V_{C_\textrm{transfer}}^\textrm{end post} &= \frac{V_{C_\textrm{causal}}^\textrm{begin post} C_\textrm{causal} }{C_\textrm{transfer}+C_\textrm{casual}}\label{eq:vtransferafterpostsimpl}
	\end{align}

The capacitance of $C_\textrm{transfer}$ is adjustable in four steps to allow the reduction of synapse-to-synapse variations.

\begin{figure}
	\center{\includegraphics[width=0.9\linewidth]{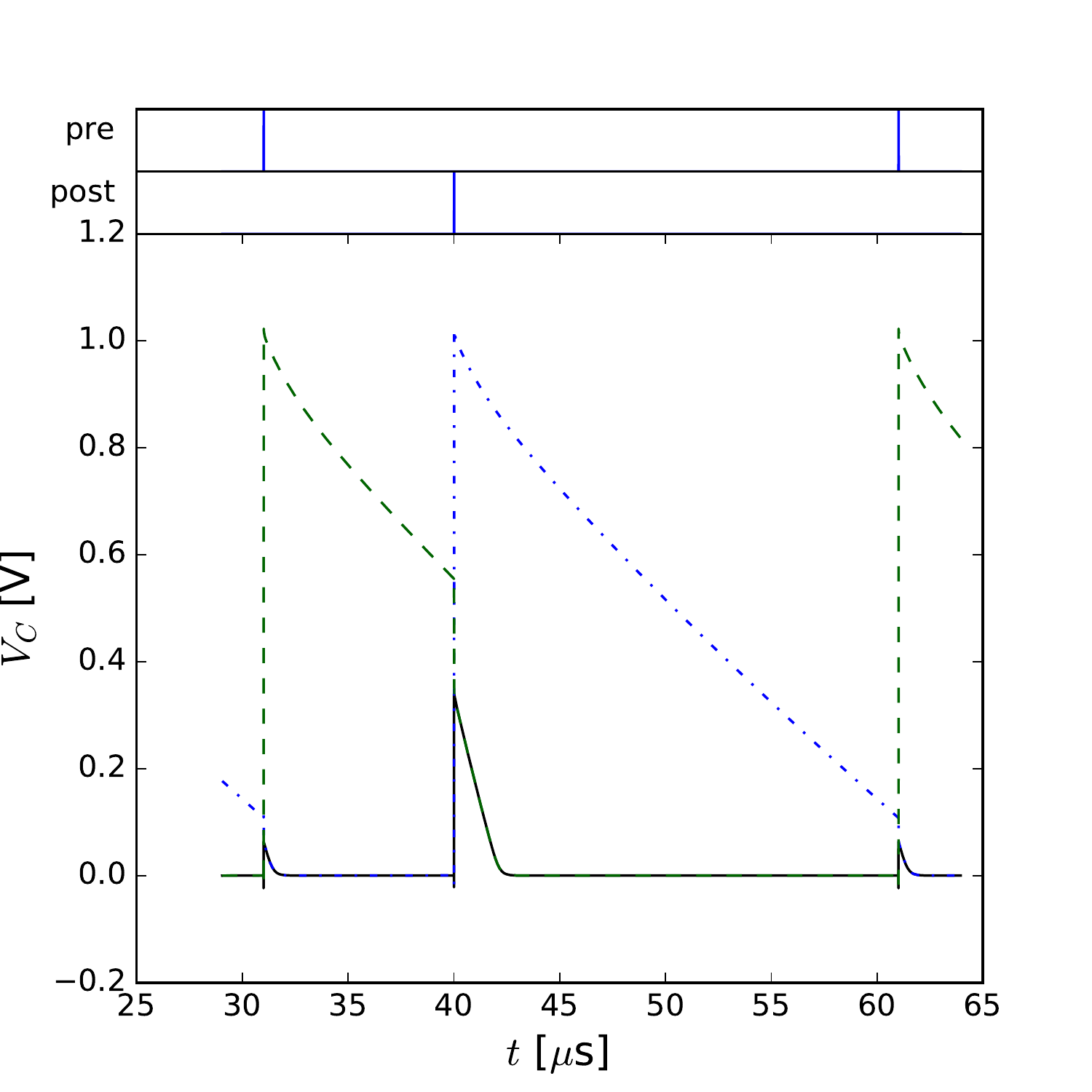}
	\caption{Simulation showing the charging of $C_\textrm{transfer}$ \added{(solid trace)} after a {\em post} pulse \added{(at $t=40\mu s$)}. \added{The dashed trace shows the voltage on $C_\textrm{causal}$ and the dashed-dotted trace the voltage on $C_\textrm{anti-causal}$.} During the charging process, the appropriate $C_\textrm{storage}$ capacitor is discharged.
	\label{fig:sim_ctransdischarge} } }
\end{figure}
Fig.~\ref{fig:sim_ctransdischarge} shows a simulation of the charging process of $C_\textrm{causal}$ and $C_\textrm{transfer}$ by M11. After the {\em post} pulse, as long as the {\em storeCausal} signal is active, $C_\textrm{causal}$ and $C_\textrm{transfer}$ are connected by M9 and their voltages are equal. The charging current is set by the gate voltage of M11. In the presented prototype chip, this voltage is directly connected to an analog input pin and set by the test controller (see Fig.~\ref{fig:overview}).

Before the time difference is stored for a causal or anti-causal measurement, its exponential value has to be calculated. This is accomplished by M13. While M13 is connected to one of the storage capacitors $C_\textrm{storage}$ by M14 or M15, it discharges the respective storage capacitor. The amount of charge it can remove from $C_\textrm{storage}$ depends on its gate voltage, which follows the time course shown in Fig.~\ref{fig:sim_ctransdischarge}.

The purpose of the charge sharing between $C_\textrm{causal}$ and $C_\textrm{transfer}$ is the reduction of the voltage representing the measured time difference below the threshold voltage of M13. This ensures the operation of M13 in weak inversion. Therefore, we can use the sub-threshold model to calculate the current through M13 at any time $t$:
\begin{align}
	I_{DS}(t) &= \frac{W}{L}I_{D0} \exp \left( \frac{V_{GS}(t)}{nkT/q} \right) \label{eq:curm13}\\
	V_{GS}(t) &=  V_{C_\textrm{transfer}}(t)\\
	I_{DS}(t) &=  I_{C_\textrm{storage}}(t)
\end{align}
Since $V_{C_\textrm{transfer}}(t)$ changes after $C_\textrm{transfer}$ has been discharged to its initial voltage during the {\em post} pulse, $V_{C_\textrm{transfer}}^\textrm{end post}$, Eq.~(\ref{eq:curm13}) has to be integrated over the time interval from the {\em post} pulse, $t_p$, to $t_e$, the point in time when $V_{C_\textrm{transfer}}(t)$ has been charged completely, i.e. $V_{C_\textrm{transfer}}(t)$ is close to zero:

\begin{align}
	\Delta Q_{C_\textrm{storage}} &= \int_{t_p}^{t_e} \! I_{C_\textrm{storage}}(t) \, \mathrm{d}t\label{eq:delqint}
\end{align}
To solve this integral a simple linear model is used for the charging of $C_\textrm{transfer}$ from $V_{C_\textrm{transfer}}^\textrm{end post}$ to zero:
\begin{align}
	V_{C_\textrm{transfer}}^\textrm{end post} &= V_{C_\textrm{transfer}}(t), t=t_p \\
	V_{C_\textrm{transfer}}(t) &=  V_{C_\textrm{transfer}}^\textrm{end post} \cdot \
	 \left( 1- \frac{t-t_p}{t_e-t_p} \
	 \right), t_p \leq t \leq t_e
\end{align}

\added{
		The time difference $t_e - t_p$ can be calculated from the current through M11 and the involved capacitances as follows:
}
\begin{align}
	t_e - t_p &=  \frac{V_{C_\textrm{transfer}}^\textrm{end post}  \cdot (C_\textrm{transfer}+C_\textrm{casual} ) } {I_\textrm{M11}}
\end{align}

Solving Eq.~(\ref{eq:delqint}) gives:
\begin{align}
	\Delta Q_{C_\textrm{storage}} &= \frac{W}{L}\frac{nkT}{q}I_{D0}\frac{t_e-t_p}{V_{C_\textrm{transfer}}(t_p)} \exp \left( \frac{V_{C_\textrm{transfer}}(t_p)}{nkT/q} -1 \right)\label{eq:delqresult}
\end{align}
Using the result of Eq.~(\ref{eq:delqresult}) the change in the voltage stored on $C_\textrm{storage}$ can be calculated:
\begin{align}
	\Delta V_{C_\textrm{storage}} &= \frac{\Delta Q_{C_\textrm{storage}} }{C_\textrm{storage}}
\end{align}
For \replaced{typical values}{the typical settings} of the transfer gain, which controls $t_e-t_p$\added{ by setting $I_{DS}$ of M11}, the deviation between Eq.~(\ref{eq:delqresult}) and the ideal exponential activation function is below \unit{1}{\%}. Also, due to the exponential decay of $I_{C_\textrm{storage}}(t)$, only the very first part of the charging of $C_\textrm{transfer}$ contributes to $\Delta V_{C_\textrm{storage}}$ significantly. If the discharge of $C_\textrm{transfer}$ is interrupted by an arriving {\em pre} pulse, the resulting error is minimal.

No control signal is needed to end the charging of $C_\textrm{transfer}$, avoiding any distortions caused by clock-feedthrough. The current $I_{C_\textrm{storage}}(t)$ is reduced to the minimum sub-threshold current without negative gate overdrive as $V_{GS_\textrm{M13}}$ approaches \unit{0}{V}. Since M13 is a thick oxide transistor with a long gate, this current is below \unit{1}{nA}. Measured total leakage on $C_\textrm{storage}$ was \unit{1.7}{mV/ms} at \unit{50}{\degree} and only \unit{0.14}{mV/ms} at room temperature (approx. \unit{25}{\degree}). The usable dynamic range of $V_{C_\textrm{storage}}$ is \unit{1.3}{V}.

M13 together with M14 or M15, respectively, also protect the thin oxide transistors used in the time difference measurement circuits from the higher supply voltage of the storage circuits. To reach sufficient storage times the utilization of thick oxide transistors is necessary to avoid gate tunneling currents. 
The gate voltage of M14 and M15 comes from the thin oxide supply voltage, thereby limiting their source voltages to save values. 

As a second function M14 and M15 act as cascodes to limit the voltage swing at the drain of M13, thereby reducing the variation of $\Delta V_{C_\textrm{storage}}$ as a function of the stored voltage on $C_\textrm{storage}$.

The storage circuits themselves use MIM-capacitors as storage cells, sitting on top of the each synapse, whereas $C_\textrm{causal}$, $C_\textrm{anti-causal}$ and $C_\textrm{transfer}$ are implemented as MOS-capacitors. $C_\textrm{transfer}$ uses several individual transistors to accomplish the digital calibration feature.

Each storage circuit uses a source follower (M18) for the readout of the stored correlation results. A pass-transistor (M19) connects the source follower to the correlation readout line if the readback enable signal of the row is active. There are two readout lines per synaptic column, thereby causal and anti-causal data of every synapse in one row can be simultaneously connected to the inputs of the correlation \gls{adc} at the top of the synapse array. 

Each storage capacitor of the synapse array can be cleared individually by activating a causal or anti-causal column correlation reset signal together with a row-wise correlation reset enable. During network operation the \gls{ppu} generates a pattern on the correlation reset inputs, depending on the results of the plasticity calculations, before it applies the column reset enable. The reset voltage can be adjusted, as can the bias current of the readback source followers, to adjust the readback voltage range to the input range of the correlation \glspl{adc}.

\subsection{Plasticity Processing Unit}
\label{sec:circ:ppu}

\begin{figure}
    \begin{center}
        \includegraphics{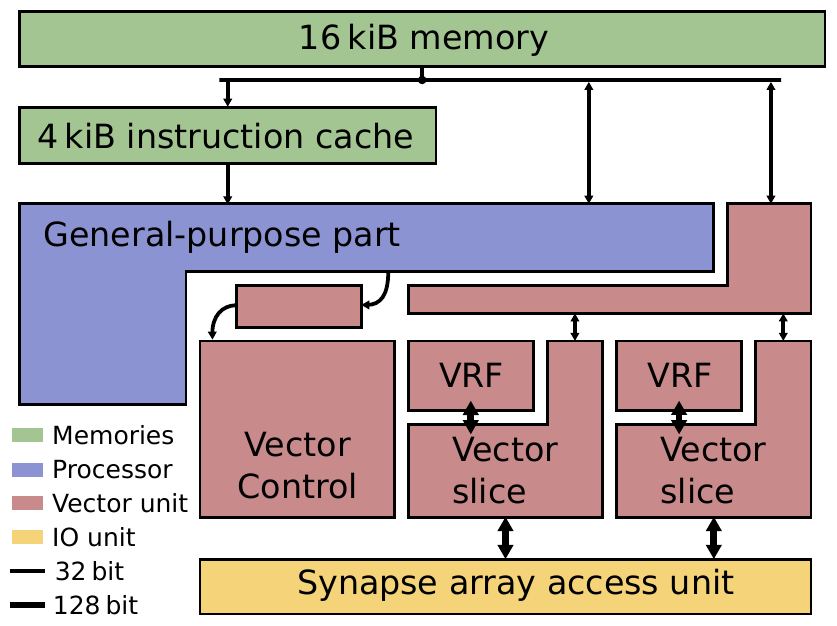}
    \end{center}
    \caption{        The \gls{ppu} is part of the plasticity sub-system and computes weight updates.
        It consists of a general-purpose part implementing the Power \gls{isa} and a special-function unit to accelerate computations using \gls{simd} operations.
        The processor has access to \unit[16]{kiB} of on-chip memory and uses a \unit[4]{kiB} direct-mapped instruction cache.
        The special-function unit consists of a shared control unit for multiple datapath slices operating on \unit[128]{bit} vectors.
        See Fig.~\ref{fig:vectorunit} for details of the vector unit.
    }
    \label{fig:ppuoverview}
\end{figure}

\begin{figure*}
    \begin{center}
        \includegraphics{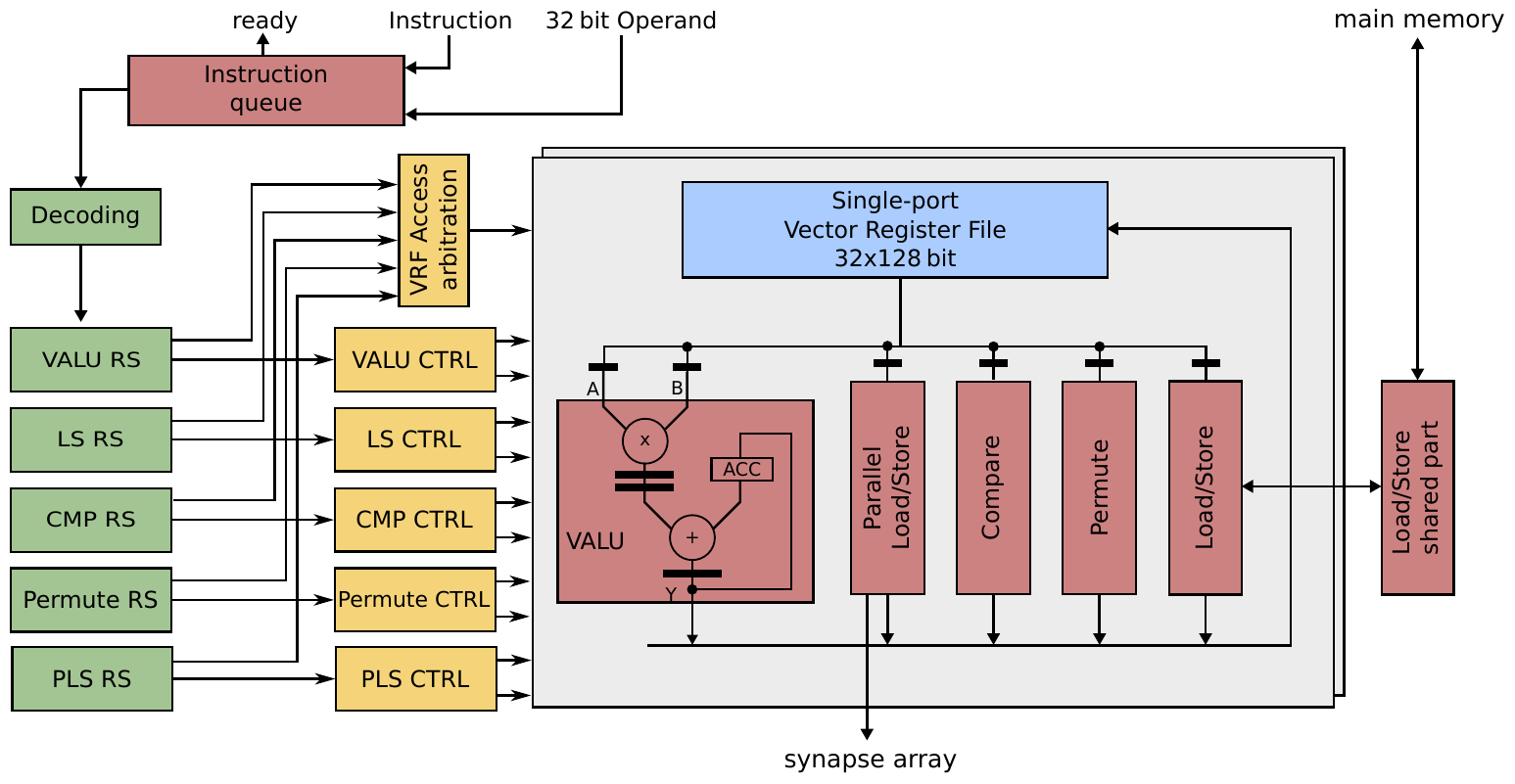}
    \end{center}
    \caption{        Detailed view of the special-function unit for \gls{simd} operations within the \gls{ppu}.
        The general-purpose part sends instructions with optionally a \unit[32]{bit} operand from the general-purpose register file via a queue.
        The decoding unit dispatches instructions to the respective reservation station upon availability.
        When operands are available and the execution unit is ready, the reservation station issues the operation to the control unit, which controls the multiple parallel datapaths.
        The vector register file has a single port for reading and writing.
        Access is arbitrated between reservation stations using a pseudo-random fair scheme.
        The serial load/store unit accesses main memory through a shared datapath.
    }
    \label{fig:vectorunit}
\end{figure*}

Fig.~\ref{fig:ppuoverview} shows an overview of the \gls{ppu}.
It is a general-purpose micro-processor extended with a functional unit specialized for parallel processing of synapses.
The general-purpose part implements the Power \gls{isa} 2.06~\cite{powerisa_206} in order to be compatible with existing compilers.
We have chosen a \unit[32]{bit} embedded implementation.
Instructions are issued in order and can retire out of order.
The core does not have a floating-point unit, but includes fixed-point hardware multiplier and divider.
In the presented chip it has access to \unit[16]{kiB} of main memory with a direct-mapped instruction cache of \unit[4]{kiB}.
The SystemVerilog source code of the implementation is available as open source from \cite{githubnux}.

The special-purpose functional unit implements an instruction set extension for accelerated processing of synapses.
Following the \gls{simd} principle a single control unit operates multiple -- two for the presented chip -- datapath slices.
Each slice operates on \unit[128]{bit} wide vectors of either eight or sixteen elements.
Of these vectors 32 can be stored in a dedicated register file in each slice.
Fig.~\ref{fig:vectorunit} shows a block diagram of the unit.

The vector unit is organized as a weakly coupled co-processor with five functional units that have their own reservation stations.
Upon encountering vector instructions, the general-purpose part sends them to a queue, which completes execution on the general-purpose side.
The vector unit takes instructions in order from this queue, decodes them and distributes them to the appropriate functional units.

\begin{table}
    \centering
    \caption{Implemented Operations}
    \label{tab:ops}
    \begin{tabular}{ll}
        \toprule
        \textbf{Category} & \textbf{Operations} \\
        \midrule
        Modular \unit[16]{bit}                & mult-acc, mult, add, sub, cmp \\
        Modular \unit[8]{bit}                 & mult-acc, mult, add, sub, cmp \\
        Saturating \unit[16]{bit} fractional  & mult-acc, mult, add, sub \\
        Saturating \unit[8]{bit} fractional   & mult-acc, mult, add, sub \\
        Permutation                           & select, shift, pack, unpack \\
        Load/store                            & load/store parallel, load/store serial \\
        \bottomrule
    \end{tabular}
\end{table}

The five functional units provide operations for arithmetics, comparison, permutation, load/store from main memory, and load/store from synapses.
Tab.~\ref{tab:ops} lists what types of operations are implemented.
All operations are available in two modes treating their operands either as vectors of sixteen \unit[8]{bit} or eight \unit[16]{bit} elements.
This allows trade-offs between throughput and accuracy and is also necessary to support the capability of combining synapses to achieve weights of higher resolution.
A minimum of \unit[8]{bit} is required, since the \gls{adc} uses that particular resolution.
In addition to the two modes of different size, vector elements can be treated either to be in signed integer or signed fractional representation.
For the latter case saturating arithmetic is used, while integers always use modular arithmetic.
The arithmetic functional unit is centered around a fused multiply-add data path, which also executes instructions for simple addition, subtraction, and multiplication.

The comparison unit writes results to a vector condition register holding flags for equality, less than, and greater than for each byte.
These flags can be used by a select operation provided by the permutation unit to selectively combine two registers into one depending on a previous compare operation.
Also, arithmetic and load/store operations support conditional execution using the vector condition register.
Further operations provided by the permutation unit are bit-shifting, loading vectors from general-purpose registers, and conversion between fractional \unit[16]{bit} and storage representation.

The two load/store units serve different purposes:
one is meant for initialization of vector registers by sequentially loading words of \unit[32]{bit} length from main memory.
The other uses a fully parallel bus for accesses on synapses and the \gls{adc}.
In the presented chip this bus has a width of \unit[256]{bit}.

\subsection{Input/Output with Analog Part}
\label{sec:ppuio}
A specialized \gls{io} unit translates the load and store operation on the parallel bus into transactions to the appropriate blocks based on the used address.
Potential targets are synapse memory, \gls{adc}, and correlation readout.
Typically, the \gls{ppu} will iterate over all rows of synapses sequentially reading weights and correlation data and writing back updated weights.
Therefore, the access unit allows multiple transactions to be in progress simultaneously. 
For example, performing a \gls{sram} read operation, while an analog-to-digital conversion of correlation data is ongoing.

The presented chip can process 32 synapses in parallel, when using byte-mode operations.
Therefore, it takes two steps to compute updates for a full row of 64 synapses.
Since \gls{io} operations work on full rows, the access unit supports buffering:
results are kept in the output registers of analog blocks after a read transaction completes.
If the next read refers to the same row, the buffered results are returned immediately.

The access unit also executes requests from outside of the chip performed through a \unit[32]{bit} wide bus.
Arbitration with \gls{ppu} accesses uses a pseudo-random fair scheme:
\added{a flip-flop indicates which requester is favored upon conflict.
For every conflict the state of the flip-flop is inverted.}

\subsection{Considerations for Plasticity Processing}
The architecture includes several design decisions geared towards the main use-case of computing weight updates.
Synaptic plasticity models from biology are typically local to the synapse, i.e.\ synapses can be computed independently.
This is true for classical \gls{stdp} models \cite{caporale08_stdp,morrison08_stdp} and many phenomenological models \cite{markram97regulation,Sjoestroem2001,Sjoestroem2004,Froemke2010,Froemke2010a}.
Therefore, parallel processing of synapses is viable and we realize this using the \gls{simd} approach.

The vector unit is weakly coupled to the general-purpose part of the processor:
the two parts do not synchronize instruction execution or share instruction tracking logic.
Only when the instruction queue is full, does the general-purpose part stall.
This allows to overlap execution in both parts to a large extent.
The general-purpose part is primarily concerned with control-flow and sends the plasticity kernel to the vector unit as a stream of instructions.

For the execution of the plasticity kernel it is important, that \gls{io} accesses and computation are pipelined to achieve good performance.
While new weights for the current row of synapses are computed, the \gls{adc} should simultaneously convert analog values for the next row.
To achieve this in an efficient and automatic way, we use reservation stations for out of order execution of vector operations.
Each functional unit shown in Fig.~\ref{fig:vectorunit} has a reservation station (shown in green).
Within one reservation station instructions are issued in order.

Implementing several reservation stations is more costly than following a simpler scheme for in-order issue as it is done in the general-purpose part.
Because control logic is shared for all vector slices, this additional cost does not impinge on scalability to larger synapse arrays.
On the other hand, area of the vector slices themselves has to be minimized.
This reflects for example in the use of a single-port register file instead of a more typical three-port variant.
Thereby, register access is a bottleneck for execution -- an instruction will typically read two operands and write one result requiring three cycles on the register file -- that has to be minimized.
Therefore, we opted for a multiply-accumulate unit with internal accumulator, so that multiplication and summation can be done in one instruction and instructions can be chained without dependency on the register file.

Apart from that, we selected a minimal set of instructions focusing on fixed-point arithmetics and \gls{io} operations to save area in the vector slices.
The only concession are pack and unpack operations as part of the permute unit to efficiently convert between weight representations for storage and computation (see Section~\ref{sec:bitrepr}).

To save power while plasticity is not needed at all or waiting for the next update cycle, the clock of the \gls{ppu} is gated.
The clock is disabled when the \gls{ppu} enters the sleep state by executing the Power \gls{isa} wait instruction.
Any interrupt request, for example from a timer or an external request, re-enables the clock and wakes the processor up.

\section{Theory and Methods}
\label{sec:theory}

\begin{figure}
    \centering
    \includegraphics{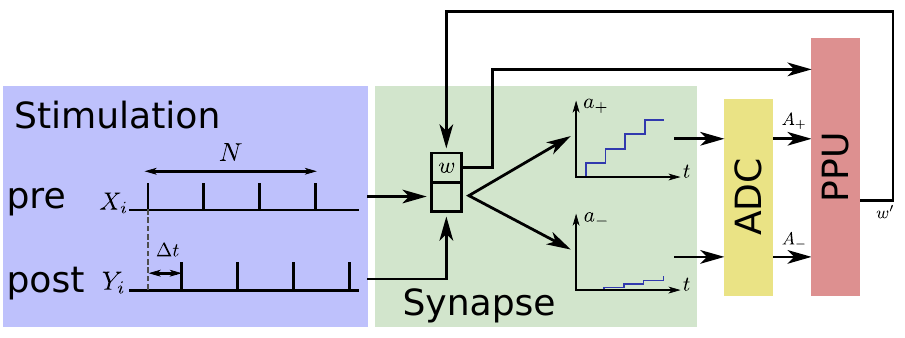}
    \caption{        Protocol for single synapse experiments.
        Regular spike trains with a relative shift of $\Delta t$ are sent to the pre and post inputs of the correlation measurement circuit in the synapse.
        The local traces $a_\pm$ are read out using the \gls{adc}.
        For experiments reported in Section~\ref{sec:full} the \gls{ppu} computes new weights.
    }
    \label{fig:stdproto}
\end{figure}

Fig.~\ref{fig:stdproto} shows the experimental protocol used for simulations with the \gls{ppu} and all later measurements in hardware.
The synapse is stimulated with presynaptic spikes at times $X_i = 0, T, 2T, \ldots, NT$ where $T$ is the interspike interval and $N$ is the total number of spikes.
Postsynaptic spikes are shifted by $\Delta t$ giving firing times $Y_i = \Delta t, T + \Delta t, 2T + \Delta t, \ldots, NT + \Delta t$.
The synapse circuit accumulates this stimulation into the two local traces $a_+$ and $a_-$ as described in Section~\ref{sec:circ:synapse}.
The \gls{adc} converts the analog traces into \unit[8]{bit} digital values $A_+$ and $A_-$, respectively.
These values together with the synaptic weight $w$ are the input for the \gls{ppu} that computes the new weight $w^\prime$.

For this study we use a multiplicative \gls{stdp} rule as reference model (see for example~\cite{morrison08_stdp}):
\begin{eqnarray}
    w^\prime = w + \begin{cases}
        \lambda (w_\text{max} - w) \exp\left( -\frac{\delta}{\tau_+} \right) & \text{for } \delta > 0 \\
        -\lambda\alpha w \exp\left( \frac{\delta}{\tau_-} \right) & \text{for } \delta \le 0
    \end{cases}
    \label{eqn:multstdp}
\end{eqnarray}
Here, $\lambda$ is a scaling parameter,  $w_\text{max}$ is the maximum weight, $\delta$ is the time difference between pre- and postsynaptic firing ($\delta > 0$ if the presynaptic event occurs before the postsynaptic one), $\tau_\pm$ are time constants, and $\alpha$ controls the asymmetry between the pre-before-post ($\delta > 0$) and post-before-pre ($\delta < 0$) branches.

The exponential term in (\ref{eqn:multstdp}) is realized by the synapse circuit itself (see Section~\ref{sec:circ:synapse}) and accumulated on the local traces $a_\pm$.
The $a_\pm$ correspond to the voltage on $C_\text{storage}$ in the synapse circuit.
We use a different symbol here to refer to the value visible to the \gls{ppu}, i.e.\ including offset from the source follower of the \replaced{readout circuit}{read out}.
The $a_\pm$ are also inverted compared to the physical voltage, so that $a_\pm = \unit[0]{V}$ corresponds to the reset value on $C_\text{storage}$.
In an idealized model of the actual circuit, these traces are given by summing over previously observed spike-pairs
\begin{eqnarray}
    a_+ &=& \sum_{\text{pre-post pairs}} \eta_+ \exp\left( -\frac{\delta}{\tau_+} \right) \label{eqn:ap} \\
    a_- &=& \sum_{\text{post-pre pairs}} \eta_- \exp\left( \frac{\delta}{\tau_-} \right) \label{eqn:am}
\end{eqnarray}
with the analog accumulation rates $\eta_\pm$.
The summed up pairs are selected according to a reduced symmetric nearest neighbor pairing rule as defined in \cite{morrison08_stdp}.
This is the same pairing scheme as was already used in \cite{schemmel_ijcnn06}.
To approximate the rule described by (\ref{eqn:multstdp}), the \gls{ppu} uses the converted digital values $A_\pm$ to compute
\begin{eqnarray}
    A &=& A_+ - A_- \label{eqn:A} \\
    w^\prime &=& w +
    \begin{cases}
        \lambda (w_\text{max} - w) A & \text{for } A > 0 \\
        \lambda\alpha w A & \text{else.}
    \end{cases} \label{eqn:wup}
\end{eqnarray}
After the update, the accumulation traces $a_\pm$ are reset to zero.

\subsection{STDP Interaction Box}
\label{sec:stdpbox}
To quantify the measured \gls{stdp} curves we extract two measures from the observed $a_\pm(\Delta t)$ dependency:
the amplitude $\hat{a}_\pm$ and the \gls{fwhm} $\hat{\tau}_\pm$.
For illustration they are plotted together as a box with height $\hat{a}_\pm$ and width $\hat{\tau}_\pm$ in Fig.~\ref{fig:variation}\deleted{D}.
The amplitude is given as
\begin{eqnarray}
    \hat{a}_\pm &=& \max{a_\pm} - \min{a_\pm}.
\end{eqnarray}
\gls{fwhm} is given as the range where $a_\pm$ is below $\frac{1}{2}\left( \max{a_\pm} - \min{a_\pm} \right) + \min{a_\pm}$.

\subsection{Bit-Representation of Weights}
\label{sec:bitrepr}

Each synapse provides \unit[6]{bit} of \gls{sram} memory for weight storage.
Two synapses can be combined to increase the effective weight to \unit[12]{bit}.
The \gls{ppu} uses either \unit[8]{bit} or \unit[16]{bit} operations giving some freedom in how weights are represented for computation.
For this study, we use a fractional number format with saturating arithmetic\added{, i.e.\ over- and underflows are prevented by saturating to maximum and minimum values \cite{liu2008numerical}.}
Weights are aligned to use the range from $0$ to $1$, i.e.\ one zero bit is added to the right for \unit[8]{bit} computations:
\begin{center}
    \begin{bytefield}[bitwidth=3em]{8}
        \bitheader[endianness=big]{7,0} \\
        \bitbox{1}{$-1$} & \bitbox{1}{$2^{-1}$} & \bitbox{1}{$2^{-2}$} & \bitbox{1}{$2^{-3}$} & \bitbox{1}{$2^{-4}$} & \bitbox{1}{$2^{-5}$} & \bitbox{1}{$2^{-6}$} & \bitbox{1}{$2^{-7}$} \\
        \bitbox{1}{0} & \bitbox{1}{$w_5$} & \bitbox{1}{$w_4$}  & \bitbox{1}{$w_3$}  & \bitbox{1}{$w_2$}  & \bitbox{1}{$w_1$} & \bitbox{1}{$w_0$} & \bitbox{1}{$0$}
    \end{bytefield}
\end{center}
Here, the $w_i$ are the individual bits of the weight with $w_5$ being the \gls{msb}.
For \unit[12]{bit} weights the representation is as follows:
\begin{center}
    \begin{bytefield}[bitwidth=3em]{8}
        \bitheader[lsb=8,endianness=big]{15,8} \\
        \bitbox{1}{$-1$} & \bitbox{1}{$2^{-1}$} & \bitbox{1}{$2^{-2}$} & \bitbox{1}{$2^{-3}$} & \bitbox{1}{$2^{-4}$} & \bitbox{1}{$2^{-5}$} & \bitbox{1}{$2^{-6}$} & \bitbox{1}{$2^{-7}$} \\
        \bitbox{1}{0} & \bitbox{1}{$w_{11}$}& \bitbox{1}{$w_{10}$}& \bitbox{1}{$w_{9}$}& \bitbox{1}{$w_{8}$}& \bitbox{1}{$w_{7}$} & \bitbox{1}{$w_6$} & \bitbox{1}{$w_5$} \\
        \bitheader[endianness=big]{7,0}\\
        \bitbox{1}{$2^{-8}$} & \bitbox{1}{$2^{-9}$} & \bitbox{1}{$2^{-10}$} & \bitbox{1}{$2^{-11}$} & \bitbox{1}{$2^{-12}$} & \bitbox{1}{$2^{-13}$} & \bitbox{1}{$2^{-14}$} & \bitbox{1}{$2^{-15}$} \\
        \bitbox{1}{$w_4$} & \bitbox{1}{$w_3$} & \bitbox{1}{$w_2$} & \bitbox{1}{$w_1$} & \bitbox{1}{$w_0$} & \bitbox{1}{$0$} & \bitbox{1}{$0$} & \bitbox{1}{$0$}
    \end{bytefield}
\end{center}
Bits $w_{11}\ldots w_{7}$ are physically stored in one synapse, while $w_6 \ldots w_0$ reside within the other one.
Special pack and unpack operations are implemented to facilitate conversion between the shown representation for computation and the stored representation.

Since for this study synaptic transmission of events to the neuron is not used, weights are permanently kept in a vector register.
So no \gls{io} operations are performed.

\section{Simulations}
\label{sec:sim}
\label{sec:sim:ppu}

\begin{figure}
    \centering
    \includegraphics{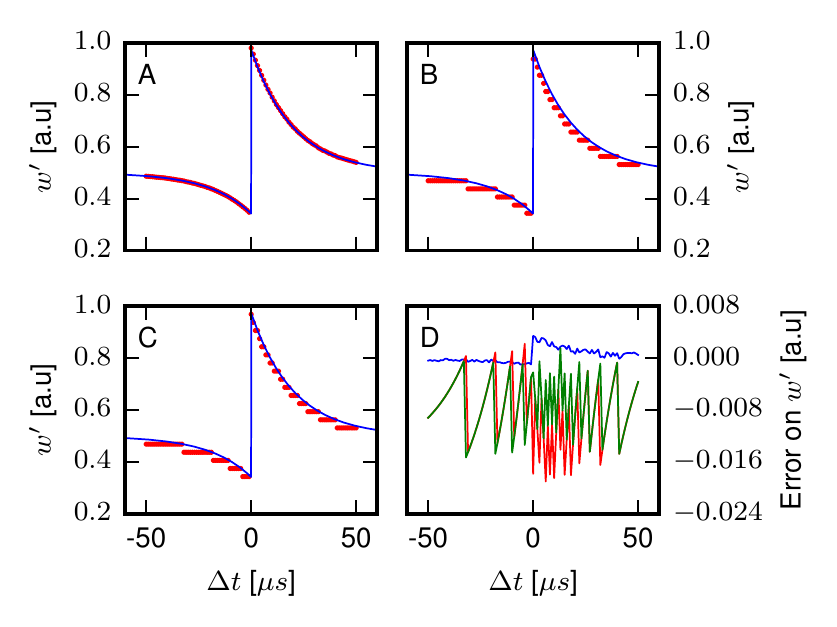}
    \caption{        Simulation results for weight \replaced{updates}{update} with idealized synapses and \gls{adc}.
        The red points show the weight $w^\prime$ as computed by the \gls{ppu} after stimulation.
        The blue lines show the result $w^\prime_\text{theo}$ with perfect precision.
        \textbf{(A)} \unit[16]{bit} computational mode for \unit[12]{bit} weight resolution.
        \textbf{(B)} \unit[8]{bit} computational mode for \unit[6]{bit} weight resolution.
        \textbf{(C)} \unit[16]{bit} computational mode for \unit[6]{bit} weight resolution.
        \textbf{(D)} Error $w^\prime - w^\prime_\text{theo}$ introduced by limited numerical precision.
        (blue: A, red: B, green: C)
    }
    \label{fig:stdpcurvesim}
	
\end{figure}

To quantify the inaccuracies added by weight resolution and numerical precision of computations performed by the \gls{ppu}, we simulate the protocol outlined in Fig.~\ref{fig:stdproto} and Section~\ref{sec:theory} with an idealized synapse circuit and \gls{adc}.
This means, that accumulation by the synapse follows \added{equations} (\ref{eqn:ap}) and (\ref{eqn:am}) exactly.
The \gls{ppu} computes weight updates according to (\ref{eqn:A}) and (\ref{eqn:wup}) using \unit[8]{bit} mode for \unit[6]{bit} weight resolution and \unit[16]{bit} mode for \unit[6]{bit} and \unit[12]{bit} weight resolutions.

\subsection{Numerical Accuracy}

Fig.~\ref{fig:stdpcurvesim} shows results for $N=32$ spike-pairs with time-constants $\tau_\pm = \unit[20]{\mu s}$ and accumulation rates $\eta_\pm = \unit[0.25]{V}$.
Weights are computed with $\lambda = 0.4$, $w_\text{max} = 1$, and $\alpha = 1$.
The initial weight is $w = 0.5$.
The blue curves show the predicted result for updates performed without limited numerical precision based on the accumulated values $a_\pm$.
The residuals shown in Fig.~\ref{fig:stdpcurvesim}D therefore represent the error introduced by discretization of $a_\pm$ to \unit[8]{bit} values $A_\pm$ in the \gls{adc} and numerical errors introduced by fixed-point arithmetic.
This error is generally small: below $3.4\times 10^{-3}$ for Fig.~\ref{fig:stdpcurvesim}A, below $1.9\times 10^{-2}$ for Fig.~\ref{fig:stdpcurvesim}B, and below $1.5\times 10^{-2}$ for Fig.~\ref{fig:stdpcurvesim}C.

Notably, \unit[6]{bit} weights systematically are smaller than predicted.
According to these results, the use of the \unit[16]{bit} mode for \unit[6]{bit} synapses reduced the error especially for large updates, i.e.\ small $|\Delta t|$.

\subsection{Updating Performance}

\begin{table}
    \centering
    \caption{Update Rates in Simulation}
    \label{tab:updaterate}
    \begin{tabular}{rcrrrrr}
        \toprule
        \textbf{No.} & \textbf{ADC} & \textbf{Mode} & \textbf{Resolution} & \textbf{Cycles} & \textbf{Row time} & \textbf{Bio. rate} \\
        \midrule
        1. & no  & \unit[8]{bit}  & \unit[6]{bit}  & 5122   & \unit[320]{ns} & \unit[97.6]{Hz} \\
        2. & no  & \unit[16]{bit} & \unit[12]{bit} & 4074   & \unit[255]{ns} & \unit[122.7]{Hz} \\
        3. & yes & \unit[8]{bit}  & \unit[6]{bit}  & 11957  & \unit[747]{ns} & \unit[41.8]{Hz} \\
        4. & yes & \unit[16]{bit} & \unit[12]{bit} & 6245   & \unit[390]{ns} & \unit[80.1]{Hz} \\
        \bottomrule
    \end{tabular}
\end{table}

The simulation used for the previous section also provides performance results in terms of achievable update rates.
Depending on the learning task a minimum update rate may be required for correct functionality \cite{Pfeil2012}.
The classical model of \gls{stdp} assumes immediate updates to the weight and so any delay can lead to mismatch to software simulations.
Tab.~\ref{tab:updaterate} shows performance results for four different scenarios with and without \gls{adc} conversions and for different weight resolutions.
The number of cycles represents the total time to update the full array of synapses.
Row time is the resulting duration for a single row assuming a clock frequency of \unit[500]{MHz}.
The biological update rate shows the frequency of updates as seen by a single synapse translated into the biological time domain.
The latter number assumes, that the update program iterates over all rows updating synapses in turn and is therefore a worst-case estimate.

The update frequencies are in all cases high compared to spike frequencies in the range of approximately \unit[1-15]{Hz} expected from biology \cite{greenberg2008population,kock2009spiking,potjans2012cell}.
A previous study has identified \unit[1]{Hz} as a lower threshold for a particular correlation detection task \cite{Pfeil2012}.
However, their updating mechanism did not use an \gls{adc} but only employed a threshold comparison leading to larger errors on the accumulation traces $a_\pm$ for longer delays.
It is, therefore, conceivable that for the same task the \gls{ppu}-based approach is less sensitive to update frequency.

The \gls{adc} requires \unit[560]{ns} for the conversion of one row of synapses.
Rows 1 and 2 in Tab.~\ref{tab:updaterate} show that all other operations can execute in less time.
Therefore, conversion by the \gls{adc} limits the update rate.
Updates for \unit[12]{bit} weights are generally faster, because two rows of synapse circuits are combined into one logical one.
This leads to half the number of \gls{adc} conversions and computational operations.
The additionally required pack and unpack operations to convert between stored and logical representation (see Section~\ref{sec:bitrepr}) do not impact performance.

\section{Experiments}
\label{sec:experiments}

Fig.~\ref{fig:board} shows the produced chip and the test setup used for experiments.
The chip contains 64 neurons with 32 synapses each for a total of 2048 synapses. 
A single-ended \gls{serdes} link provides communication with a Xilinx Spartan-6 \gls{fpga} for control and event data.
Link and internal logic operate with the same clock signal provided via a chip pin.
The system is designed for frequencies up to \unit[500]{MHz} and operated at \unit[97.5]{MHz} in this study.

The \gls{fpga} is equipped with \unit[512]{MiB} of DDR3-SDRAM and communicates with a PC via USB~2.0.
Due to the real-time nature of neuromorphic hardware and the small time-scales involved, communication with the chip is buffered in the on-board SDRAM attached to the \gls{fpga} and played-back under precise timing control.
The \gls{fpga} uses a byte-code with instructions of variable length to provide efficient coding with \unit[64]{bit} effective time stamp resolution.
The byte-code is executed at a clock frequency of \unit[97.5]{MHz} leading to a best-case temporal precision of \unit[10.26]{ns}.
Responses and events are recorded with annotated timing information using the same byte-code representation.

\subsection{Weight Linearity}

\begin{figure}
    \centering
    \includegraphics{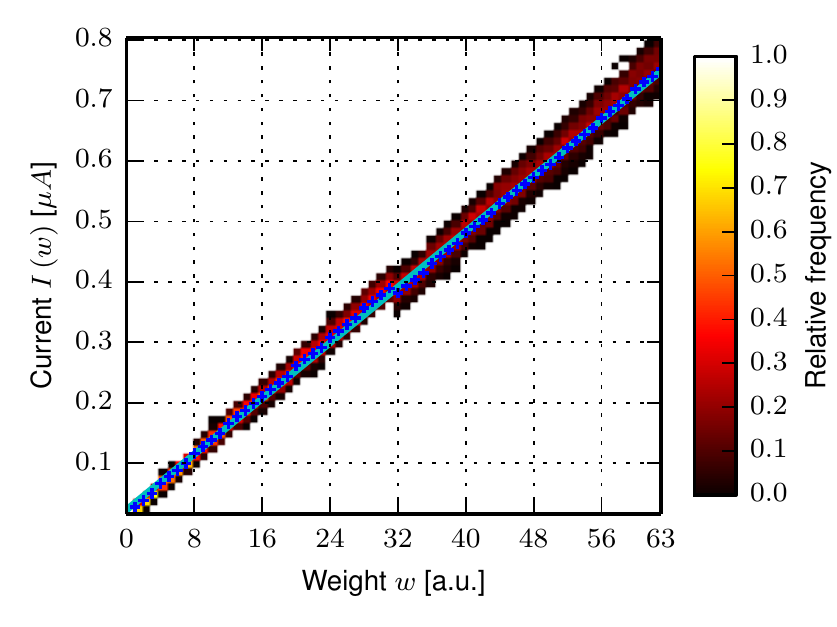}
    \caption{        Output current from the \gls{dac} within the synapse.
        The measurement includes 96~synapses (three columns) from one chip.
        Blue crosses mark the mean values.
        The best fit to all data points is shown as cyan colored line.
    }
    \label{fig:weight}
\end{figure}
We first analyze the \gls{dac} within the synapse.
Fig.~\ref{fig:weight} shows the average output current for a total of 96 synapses on one chip over the full range of 64 possible weight values.
The current was measured by sending a high-frequency input spike train to the synapse and measuring the resulting current using a readout pin and an external current measurement device\footnote{Keithley SourceMeter 2635}.

The fit yields an offset of $\unit[\dacOffsetP]{nA}$ and a value of $\unit[\dacSlopeP]{nA}$ for one \gls{lsb}.
With these values the maximal \gls{inl} is \unit[\dacInlP]{LSB}, while the mean \gls{inl} is \unit[\dacMeanInlP]{LSB}. The systematic shift at the transition from code 31 to 32 is caused by well-proximity effects. Two fingers of the MSB transistor of the \gls{dac} are too closed to an adjacent well. This was only discovered after tape-out.

\subsection{Variability}
\label{sec:var}

\begin{figure}
    \centering
    \includegraphics{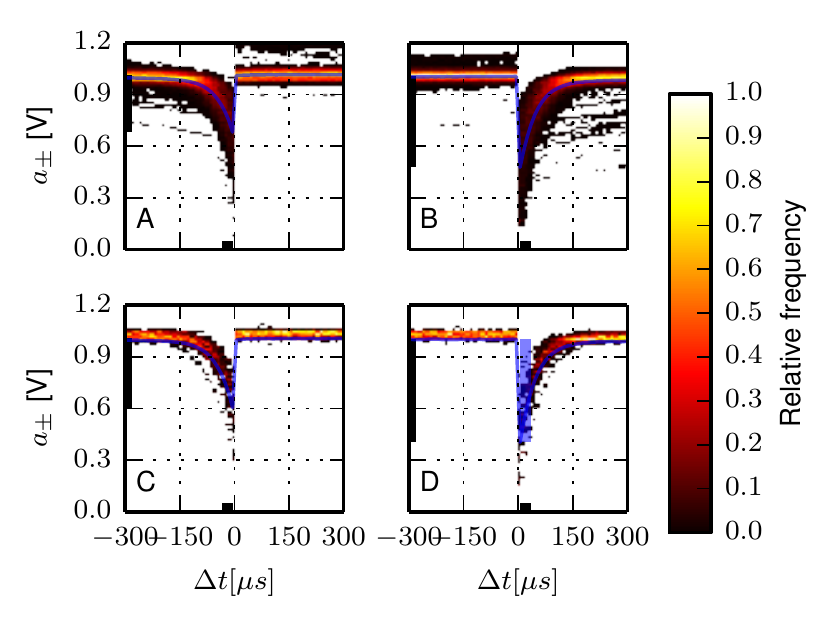}
    \caption{        Accumulation values $a_\pm$ after stimulation with $N=32$ spike-pairs as two-dimensional histogram.
        Color indicates the relative frequency.
        The mean values are plotted as blue lines.
        The black bars on the axes indicate width and amplitude of the \gls{stdp} interaction box (see Section~\ref{sec:stdpbox}), which is also shown in blue in the last picture.
        Data points are shifted to an offset without stimulation of \unit[\variationOffsetP]{V} to correct for different offsets on different chips.
        \textbf{(A)} Post-before-pre measurement $a_-$ for \variationNumSynapsesInAP synapses on three different chips.
        \textbf{(B)} Pre-before-post measurement $a_+$ for \variationNumSynapsesInBP synapses on three different chips.
        \textbf{(C,D)} Data from 32 synapses on the same \gls{adc} channel on one chip.
    }
    \label{fig:variation}
\end{figure}

Fig.~\ref{fig:variation} shows the measured dependency $a_\pm(\Delta t)$ using the experimental protocol illustrated in Fig.~\ref{fig:stdproto}.
The curves were measured using $N = 32$ spike pairs and analog parameters $V_\text{ramp} = \unit[250]{mV}$ and $V_\text{store} = \unit[350]{mV}$.
For all following experiments shown in this study ambient temperature was kept at \unit[25]{$\degree C$}.
The data shown in Fig.~\ref{fig:variation} is corrected for different offsets of the readout on different chips.
All curves are shifted vertically, so that without stimulation the average $\left< a_\pm \right>$ lies at \unit[\variationOffsetP]{V}.
This way the curves can be shown and compared in one plot.
For learning applications the offset is determined on program startup by the \gls{ppu}.

The results show biologically realistic time-constants of approximately \unit[20]{$\mu s$} to be achievable.
Here, we use a speed-up factor of $10^3$ to convert from biological time-constants of approximately \unit[20]{ms} given in \cite{markram1995action,poo98stdp,dan04stdp}.
The average time-constants in Fig.~\ref{fig:variation} are $\left<\hat{\tau}_\pm\right> = \unit[30]{\mu s}$ with a standard deviation of \unit[10]{$\mu s$} for Fig.~\ref{fig:variation}A,B and \unit[8]{$\mu s$} for Fig.~\ref{fig:variation}C,D.
The achievable ranges are discussed later (see Fig.~\ref{fig:ranges}).

Trial-to-trial variability for individual synapses is generally small.
The mean trial-to-trial standard deviation for all four plots is equal within errors at $\unit[8 \pm 5]{mV}$.
Therefore, the variation between synapses that can be seen in the plots is due to device mismatch within the synapse circuit itself and mismatch within the readout channels of the \gls{adc}.
Plots C and D of Fig.~\ref{fig:variation} show only data for a single channel each.
Concerning amplitude, standard deviations for the multi- and single-channel cases are comparable:
$\left<\hat{a}_\pm\right> = \unit[400 \pm 140]{mV}$ for A and C, $\left<\hat{a}_\pm\right> = \unit[600 \pm 180]{mV}$ for B and D.
For the time-constants single-channel data exhibit slighlty less variability (see \replaced{Fig.\ref{fig:variation} C and D}{above}).
However, differences are small and overall variability can be assumed to be dominated by mismatch between the synapse circuits themselves.

\subsection{Achievable Ranges}

\begin{figure}
    \centering
    \includegraphics{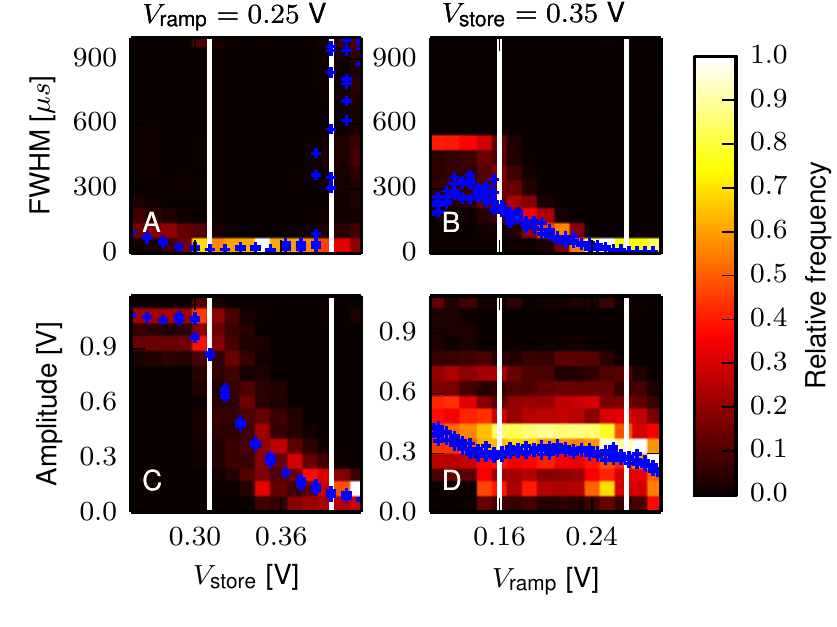}
    \caption{        Width and amplitude of the \gls{stdp} interaction box in dependence of parameters $V_\text{store}$ \added{(left column)} and $V_\text{ramp}$ \added{(right column)}.
        The blue crosses show data for one single synapse.
        White bars mark the useful range for the corresponding parameter (see text).
        Colors indicate the relative frequency for a total of 192 synapses on three different chips.
    }
    \label{fig:ranges}
\end{figure}

To configure the shape of the \gls{stdp} curve the circuit provides two primary configuration parameters: $V_\text{store}$ and $V_\text{ramp}$ (see Section~\ref{sec:circ:synapse}). \added{$V_\text{store}$ controls the storage gain and $V_\text{ramp}$ the time constant (see Fig.~\ref{fig:corrsense_circuit}).}
We measured 192 synapses on three different chips sweeping both parameters to find the achievable amplitudes $\hat{a}_\pm$ and widths $\hat{\tau}_\pm$.
Fig.~\ref{fig:ranges} shows the results while using $N = 32$ spike pairs
\added{(see Section~{\ref{sec:stdpbox} for the definiton of the plotted
quantities `width' and `amplitude'})}.

The usable range is the parameter range, for which $V_\text{store}$ controls amplitude and $V_\text{ramp}$ controls the width.
The respective other property, i.e.\ width for $V_\text{store}$ and amplitude for $V_\text{ramp}$, remains flat.
Therefore, the shape of the \gls{stdp} curve can be tuned with the given parameters.
For the presented measurements the usable ranges were selected as $V_\text{store} \in \left[\unit[\usableVstoreMinP \ldots \usableVstoreMaxP]{V}\right]$ and $V_\text{ramp} \in \left[\unit[\usableVrampMinP \ldots \usableVrampMaxP]{V}\right]$.
This range lies between the white vertical markers in Fig.~\ref{fig:ranges}.
Tab.~\ref{tab:ranges} gives mean and standard deviation at start and stop of this range for $\hat{a}$ and $\hat{\tau}$.
The amplitude covers nearly the \unit[1]{V} of full dynamic range of the \gls{adc} input.
Time-constants show a large configurable range from tens to hundreds of micro-seconds.
Even lower values down to $\unit[2]{\mu s}$ are configurable, but the error will stay at $\unit[4]{\mu s}$ so that we have excluded these values from the usable range.
The amplitude can maximally be as large as the available input range of the \gls{adc}, which is evident in the measured data.

\begin{table}
    \centering
    \caption{Achievable Ranges}
    \label{tab:ranges}
    \begin{tabular}{cccc}
        \toprule
        \textbf{Parameter} & \textbf{Start} & \textbf{Stop} & \textbf{Unit} \\
        \midrule
        $\hat{a}_+$     & $\usableVstoreMinAmplCausalMeanP \pm \usableVstoreMinAmplCausalErrP$    & $\usableVstoreMaxAmplCausalMeanP \pm \usableVstoreMaxAmplCausalErrP$ & \unit{V} \\
        $\hat{\tau}_+$  & $\usableVrampMinFwhmCausalMeanP \pm \usableVrampMinFwhmCausalErrP$    & $\usableVrampMaxFwhmCausalMeanP \pm \usableVrampMaxFwhmCausalErrP$   & \unit{$\mu s$} \\
        $\hat{a}_-$     & $\usableVstoreMinAmplMeanP \pm \usableVstoreMinAmplErrP$    & $\usableVstoreMaxAmplMeanP \pm \usableVstoreMaxAmplErrP$ & \unit{V} \\
        $\hat{\tau}_-$  & $\usableVrampMinFwhmMeanP \pm \usableVrampMinFwhmErrP$    & $\usableVrampMaxFwhmMeanP \pm \usableVrampMaxFwhmErrP$   & \unit{$\mu s$} \\
        \bottomrule
    \end{tabular}
\end{table}

\subsection{Full-System Experiments}
\label{sec:full}

\begin{figure*}
    \centering
    \includegraphics{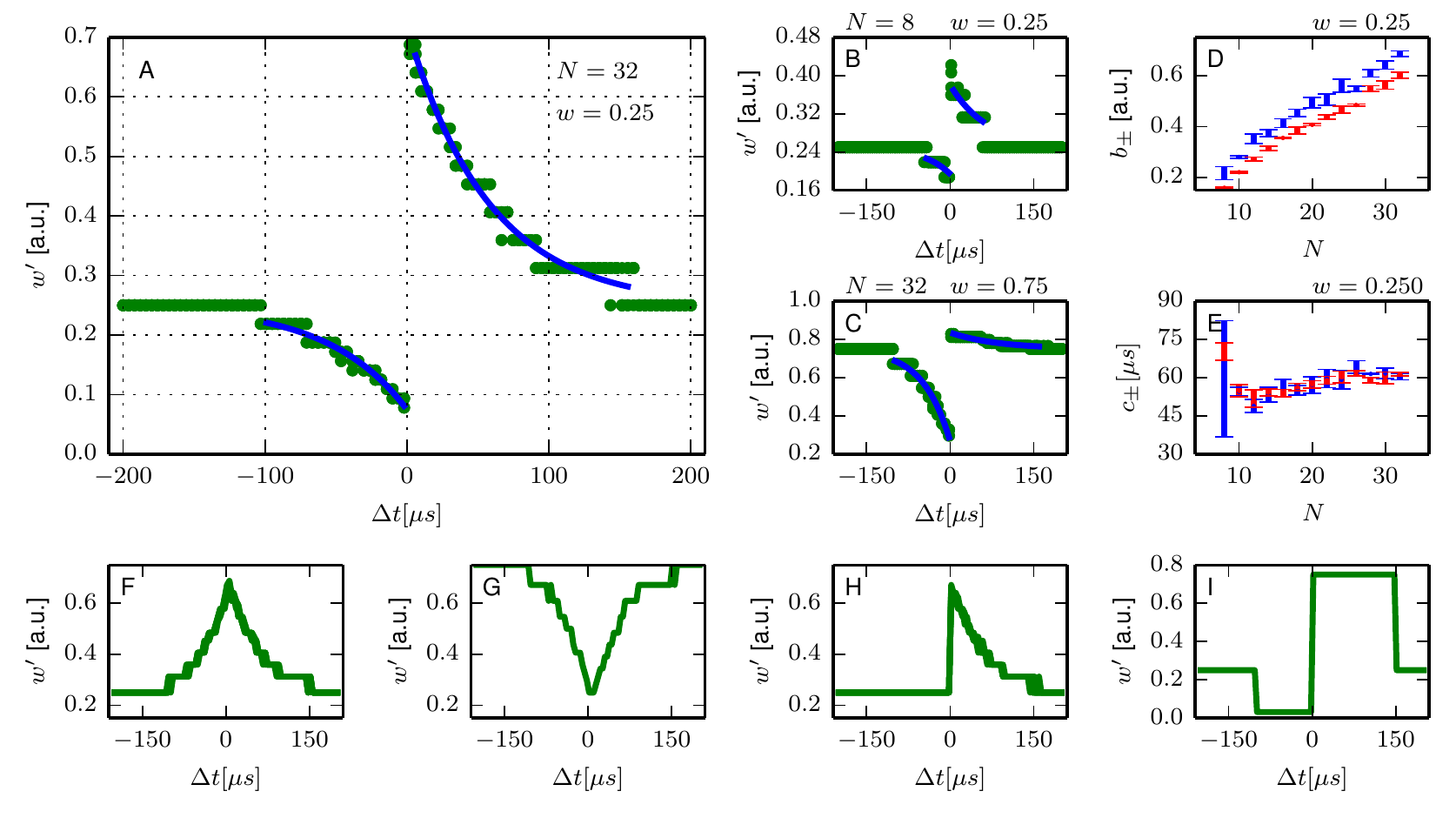}
    \caption{        Results from experiments using the full signal chain including \gls{ppu}, \gls{adc}, and correlation sensor in the synapse.
        Synapses are stimulated according to the protocol outlined in Fig.~\ref{fig:stdproto}.
        The \gls{ppu} computes a multiplicative update rule according to (\ref{eqn:Aexp}) and (\ref{eqn:wup}).
        \textbf{(A,B,C)} Weight after stimulation for five repetitions as green points.
        Fit to data as blue curve.
        \textbf{(D,E)} Resulting fit parameters for amplitude $b_+$ (red), $b_-$ (blue) and time-constant $c_+$ (red), $c_-$ (blue) for multiple number of spike-pairs $N$.
        \textbf{(F-I)} Examples of other updating rules that can be implemented.
    }
    \label{fig:ppumeasured}
\end{figure*}

With the individual channels characterized, the next step is to look at the full signal processing chain.
We use the experimental protocol described in Section~\ref{sec:theory} and illustrated in Fig.~\ref{fig:stdproto}.
The \gls{ppu} performs weight updates according to (\ref{eqn:wup}).
To eliminate trial-to-trial noise on the analog readout and to remove systematic offset between the two channels of one synapse, we modify (\ref{eqn:A}) to
\begin{eqnarray}
    \bar{A} &=& \left( A_+ - A_- \right) - A_\text{off} \label{eqn:Aoff} \\
    A &=& \begin{cases}
        \bar{A}     & \text{if } \left| \bar{A} \right| > \theta \\
        0           & \text{else}
    \end{cases} \label{eqn:Aexp}
\end{eqnarray}
Here, $A_\text{off}$ is determined at program startup after reset of the accumulation storage as difference $A_+ - A_-$.
(\ref{eqn:Aexp}) implements thresholding using the user selected parameter $\theta$.
The \gls{ppu} performs updates at regular intervals of $\unit[10]{\mu s}$ during stimulation.
The source code for the actually used update program is available from \cite{githubppusoftware}.

Fig.~\ref{fig:ppumeasured} shows results when using \unit[8]{bit} resolution for arithmetics.
For analysis, two functions $f_+$ and $f_-$ are individually fitted to pre-before-post ($\Delta t > 0$) and post-before-pre ($\Delta t < 0$) data:
\begin{eqnarray}
    f_+\left(\Delta t\right) &=& w+ b_+\left(w_\text{max}-w\right)\exp\left(-\frac{\Delta t}{c_+}\right) \\
    f_-\left(\Delta t\right) &=& w- w b_-\exp\left(\frac{\Delta t}{c_-}\right)
\end{eqnarray}
Here, $b_\pm$ and $c_\pm$ are the fit parameters, while initial weight $w$ and maximum weight $w_\text{max}$ are the same as those used by the update program.
In all experiments we set $w_\text{max}$, $\lambda$, and $\alpha$ to $1.0$.
The threshold $\theta$ was set to $10$.
Since discretization of the weight removes the long tail of the exponential, the fit is restricted to points where the weight was actually changed ($w^\prime \neq w$).

Fig.~\ref{fig:ppumeasured}A-C demonstrate different combinations of $N$ and $w$.
Especially for small updates, the discretization of the weight to \unit[6]{bit} is apparent.
Results exhibit the expected dependency on $w$ for a multiplicative rule.
Fig.~\ref{fig:ppumeasured}D,E plot the fitted parameters for amplitude ($b_\pm$) and time-constant ($c_\pm$) over the number of spike pairs $N$.
As expected, amplitude increases linearly with the number of pairs and for the chosen initial weight $w = 0.25$ positive changes are larger than negative ones.
The process that measures the timing of spike pairs in the synapse operates on individual pairs and is therefore independent of $N$.
Also, the circuit for time measurement is shared for pre-before-post and post-before-pre pairs within one synapse.
Therefore, time-constants should be identical on both sides and independent of $N$.
Experimental data is compatible with these expectations as Fig.~\ref{fig:ppumeasured}E shows.
For small $N$ the fit is not reliable due to discretization (see Fig.~\ref{fig:ppumeasured}B).

The plots in Fig.~\ref{fig:ppumeasured}F-I give a hint of the achievable flexibility.
They were produced with the same stimulation protocol only by changes in software running on the \gls{ppu}.
Fig.~\ref{fig:ppumeasured}F,G show symmetrical Hebbian and anti-Hebbian rules.
Fig.~\ref{fig:ppumeasured}H is only sensitive to pre-before-post pairings.
Fig.~\ref{fig:ppumeasured}I realizes bi-stable learning.

\subsection{Power Consumption}

During execution of the experiment described in the previous section, digital logic consumes below \unit[32]{mW} of power as measured on the power supply pins of the chip.
With the clock disabled for the \gls{ppu}, power consumption drops below \unit[10]{mW}, so that \unit[22]{mW} can be attributed to the \gls{ppu}.
In reset\added{,} power consumption drops by \unit[2]{mW} for the \gls{ppu}.
Therefore, power consumption is largely due to clock distribution.

\section{Discussion of Results}
\label{sec:discussion}

The two overarching goals in the development of neuromorphic hardware are to provide a platform for neuroscientific experiments and to find new ways of computation for technical applications.
For both these goals we believe reliability, scalability, and flexibility to be enabling factors besides efficiency in terms of power, area, and speed.
Therefore, the presented results focus on these aspects.

\subsection{Reliability}
To assess reliability we characterized the synapse behavior across three different chips (see Fig.~\ref{fig:variation} and~\ref{fig:ranges}).
Results show substantial variation due to device mismatch within the analog circuits.
Please note, that for these measurements the configuration bits of the synapse circuit were not even used (see Section~\ref{sec:circ:synapse}).
So there is room for improvements through calibration.
On the other hand, trial-to-trial variability of individual components is small.
This is for example illustrated in Fig.~\ref{fig:ranges} that shows multiple trials from a single synapse on the background of the overall distribution of synapses.
A small trial-to-trial variability was also measured for individual channels in Section~\ref{sec:var}.
This allows on the one hand to use off-line calibration, but on the other hand it is also conceivable, that an emulated network calibrates itself through the use of plasticity.
Indeed, the robustness of reward-based learning to device mismatch on the correlation detection within the processor-based approach presented here has been shown in previous work~\cite{Friedmann2013}.
Self-tuning has also been shown to be feasible through the use of short-term plasticity  \cite{bill2010compensating}.

In general, a plasticity mechanism can compensate inhomogeneities if there is a feedback loop for the parameter subject to variation.
This is typically the case for outputs, e.g.\ synaptic weights.
An \gls{stdp} rule will modify the weight according to the timing behavior it observes, which is an effect of the weight including variation.
Variation on the input however - in this case the signal $a_\pm$ from the correlation sensor - is invisible from the rule.
It can however be compensated by introducing additional information about the behavior of the system, for example through a reward signal.
This addition of complementary information is why reward-based learning rules are well suited for analog neuromorphic hardware systems.
The alternative is to use redundancy of analog components so that the average behavior is reliable.

\subsection{Scalability}
Scalability can of course only be shown by actually scaling the system, which we plan to do in the future.
Nonetheless, the plasticity system is designed to scale well:
the only part for which area scales linearly with the number of synapses is the correlation sensor that resides within the synapse circuit.
Therefore, we have chosen to use an area-optimized circuit realized as analog full-custom design.
The \gls{adc} scales with the number of columns in the synapse array, which have typically a square root dependency on the number of synapses.
Most parts of the \gls{ppu} are required only once per array and only the number of vector slices scales with the number of columns.
To keep these slices as lightweight as possible, all control logic is shared and a single-port vector register file is used.
A scaled system will feature arrays of $256 \times 256$ synapses with a dedicated \gls{ppu} using 8 vector slices.

\subsection{Flexibility}
The whole approach presented here has a strong emphasis on flexibility\added{, compared to our previous implementation \cite{schemmel_ijcnn06} and considering the constraints of an analog, accelerated neuromorphic system}.
By this we mean, that a large number of plasticity rules should be implementable in the hardware system.
Introducing the \gls{ppu} sacrifices area and power in order to have as much freedom as possible while not sacrificing speed.
To achieve this latter point in the \unit[65]{nm} technology, we consider a combination of analog and software-based processing, as shown in this study, to be necessary.
At a speed-up factor of $10^3$ and array sizes of 65k synapses it is not feasible to process individual spike events in software.
This of course limits flexibility as the functionality of the correlation sensor is fixed in hardware.
Therefore, this functionality should at least operate over a wide range of parameters\replaced{, demonstrated by the results shown in Fig.~\ref{fig:ranges} and Tab.~\ref{tab:ranges}}{. That this is possible, is shown in Fig.~\ref{fig:ranges} and Tab.~\ref{tab:ranges}.}
In the biological time domain, the design covers ranges from tens to hundreds of milliseconds, fitting typical ranges found in biology \cite{markram1995action,poo98stdp,dan04stdp}.
Also the amplitude is tunable over a large range, so that the sensitivity of the correlation sensor can be matched to the network activity.

In general, every plasticity model is implementable in this system that depends only on observables visible to the \gls{ppu} and affects only parameters accessible by the \gls{ppu}.
Observables are the weight $w$, the correlation signals $a_\pm$, a firing rate sensor not discussed in this study, and signals from outside the chip such as reward.
All parameters of the chip that can be modified at all, can also be modified by the \gls{ppu}.
This includes the synaptic weight $w$, neuron parameters, and the topology of the network.
The latter is limited to the addresses stored in the synapses for this prototype chip. 

\added{In future realizations it is feasible to increase the number of observables of the \gls{ppu}. It is planned to include a fast \gls{adc} in a forthcoming chip which will give the \gls{ppu} access to membrane voltages. It is also feasible to add synapse correlation measurement circuits with novel properties, if there are plasticity models demanding them.}

Here we only show the simple \gls{stdp} rule given in (\ref{eqn:multstdp}) as proof of concept.
Fig.~\ref{fig:ppumeasured}F-I show simple examples of modifications of the plasticity model purely realized in software running on the \gls{ppu}.
Beyond that, the reward-based learning rule R-STDP and a learning rule for spike-based expectation maximization has been ported to the system, but not yet tested in hardware \cite{Friedmann2013,breitwieser2015masterthesis}.

\section{Conclusion}
\label{sec:conclusion}

In this study we have presented a new approach to plasticity in neuromorphic hardware:
the combination of dedicated analog circuits in every synapse with a shared digital processor.
It represents a trade-off between flexibility of implementable plasticity models and efficiency of the implementation in terms of area, speed, and energy.
The presented results demonstrate the viability of this approach for plasticity.

The more classical approach taken for neuromorphic hardware, for example by \cite{qiao2015reconfigurable} or \cite{ramakrishnan2011floating}, is to implement a single plasticity mechanism that can be used to solve a range of network learning tasks. \added{Analog continuous-time implementations of neuromorphic circuits can be combined with floating-gate technology to achieve persistence of the learned synaptic weights. By modifing the control signals the precise learning rules can also be tuned \cite{brink2013}.}
Our approach not only aims for flexibility in the learning task, but also in the mechanism itself.
Together with the speed-up factor this enables experimental analysis of long-term effects of such mechanisms.
In the classical approach it is essential to have a detailed understanding of the mechanism prior to production of hardware.
In our approach the hardware system can help to gain this understanding.
This is an important aspect when designing a system intended as a neuroscientific platform.

In \cite{galluppi2015framework} this approach is taken even further:
neuronal dynamics as well as detection of correlations and weight update are performed by general-purpose processors in software.
Specialized hardware is only used for event communication.
This maximizes flexibility but further sacrifices efficiency, so that operation is only possible without speed-up. \added{Another mixed-mode approach is reported in \cite{azghadi2015programmable}. Here, the authors also perform the full plasticity operation in software, achieving maximum flexibility, while the synapses and neurons are full-custom analog implemetations.}

Our approach to use dedicated hardware for the most expensive part -- the processing of spikes -- enables faster operation. \added{Using an on-die PPU local to the synapse circuits also facilitates scaling of the system, since no communication to off-chip components is necessary.}
Since learning and development in biology are processes spanning many time-scales, platforms for accelerated simulation or emulation are important.
In the domain of general-purpose computers using software simulations even for medium-sized networks accelerated operation with plasticity is currently not possible \cite{zenke2014auryn}.

\section{Outlook}
\label{sec:outlook}

The chip presented here is still an early prototype that for example lacks on-chip networking capabilities.
However, using the experimental setup described here a wide range of plasticity mechanisms can already be implemented and analyzed in hardware.
Obvious candidates are the models already prepared for implementation \cite{Friedmann2013,breitwieser2015masterthesis}.
Future prototypes will add the ability to include neuronal and structural plasticity opening the door for a large set of learning mechanisms.
It will then also be possible to execute learning tasks involving networks of neurons with the system.

In the long run, the focus will be on scaling the system in size.
As an intermediate step we plan to build chip-scale variants with two $256 \times 256$ synapse arrays and two \glspl{ppu}.
Eventually, the goal is to go to wafer-scale \cite{schemmel2010iscas}.
It will then replace the first generation of the neuromorphic platform (NM-PM-1) of the Human Brain Project \cite{markram2011introducing}.

We also hope that the release of the \gls{ppu} design -- the Nux processor \cite{githubnux} -- as open source will turn out to be a valuable contribution to open source hardware.

\section{Acknowledgements}
This work has received funding from the European Union Seventh Framework Programme ([FP7/2007-2013]) under grant agreement no 604102 (HBP), 269921 (BrainScaleS) and 243914 (Brain-i-Nets).

\begin{IEEEbiography}[{\includegraphics[width=6pc,height=7.5pc,clip,keepaspectratio]{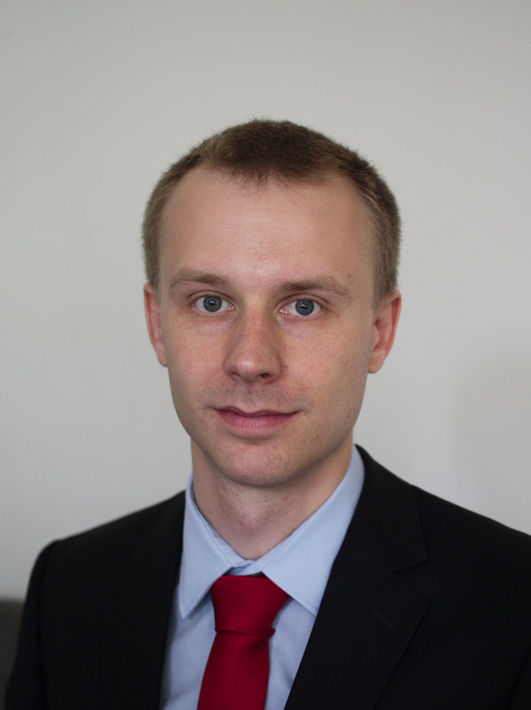}}]{Simon Friedmann}
    received the Dipl.\ phys.\ and Ph.D.\ degrees from Heidelberg University, Germany in 2009 and 2013, respectively.

    He is a post-doctoral researcher at the Electronic Vision(s) group at Heidelberg University.
    His research focus is on learning in neuromorphic hardware.
\end{IEEEbiography}
\begin{IEEEbiography}[{\includegraphics[width=6pc,height=7.5pc,clip,keepaspectratio]{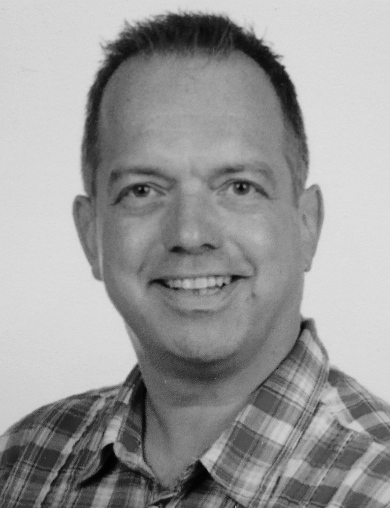}}]{Johannes Schemmel} received a Ph.D.\ in Physics in 1999 from Ruprecht-Karls University Heidelberg. He is now 'Akademischer Oberrat' at the Kirchhoff Institute of Physics in Heidelberg where he is head of the ASIC laboratory and the 'Electronic Vision(s)' research group. His research interests are mixed-mode VLSI systems for information processing, especially the analog implementation of biologically realistic neural network models. He is the architect of the Spikey and BrainScaleS accelerated Neuromorphic hardware systems.
\end{IEEEbiography}
\begin{IEEEbiography}[{\includegraphics[width=6pc,height=7.5pc,clip,keepaspectratio]{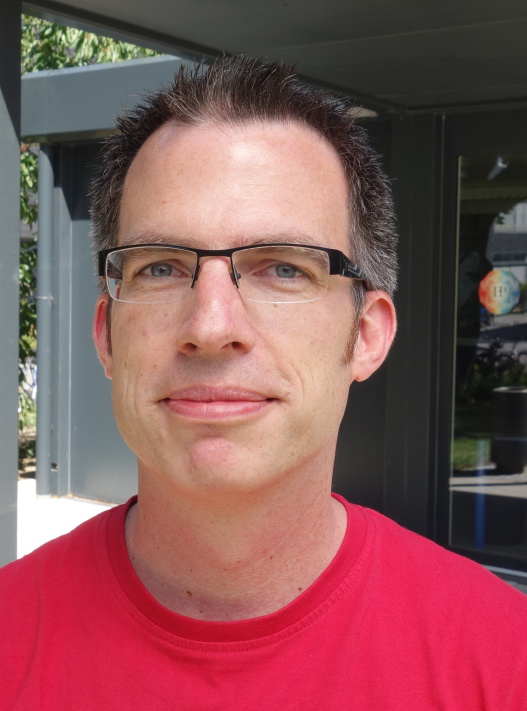}}]{Andreas Gr\"ubl} received the Dipl.\ phys.\ and Ph.D.\ degrees from Heidelberg University, Germany in 2003 and 2007, respectively. He is a senior post-doctoral researcher at the Electronic Vision(s) group and leader of the Electronics department of the Kirchhoff Institute for Physics at Heidelberg University. He has 8 years of post-doctoral experience in designing and building complex microelectronics systems for brain-inspired information processing. His research focus is on new methods for the implementation of large mixed-signal SoCs.
\end{IEEEbiography}
\begin{IEEEbiography}[{\includegraphics[width=6pc,height=7.5pc,clip,
    keepaspectratio]{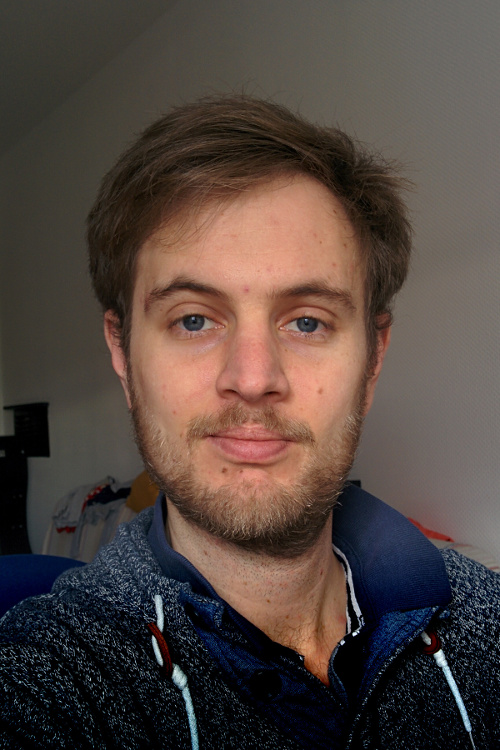}}]
    {Andreas Hartel}
    received the Dipl.\ phys.\ and Ph.D.\ degrees from Heidelberg University,
    Germany in 2010 and 2016, respectively.
    He is a post-doctoral researcher at the Electronic Vision(s) group at
    Heidelberg University. His research focus is on learning in neuromorphic
    hardware.
\end{IEEEbiography}

\begin{IEEEbiography}[{\includegraphics[width=6pc,height=7.5pc,clip,
    keepaspectratio]{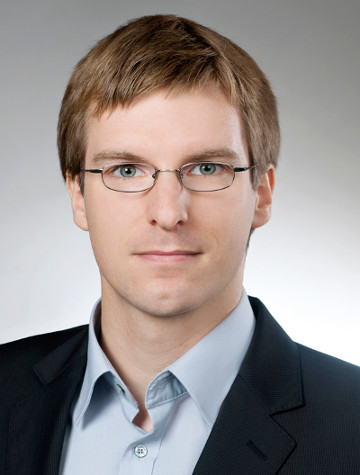}}]
    {Matthias Hock}
    received the Dipl.\ phys.\ and Ph.D.\ degrees from Heidelberg University,
    Germany in 2009 and 2015, respectively.
    He is a post-doctoral researcher at the Electronic Vision(s) group at
    Heidelberg University. His research focus is on development and test of
    mixed-signal circuits for neuromorphic hardware.
\end{IEEEbiography}

\begin{IEEEbiography}[{\includegraphics[width=6pc,height=7.5pc,clip,keepaspectratio]{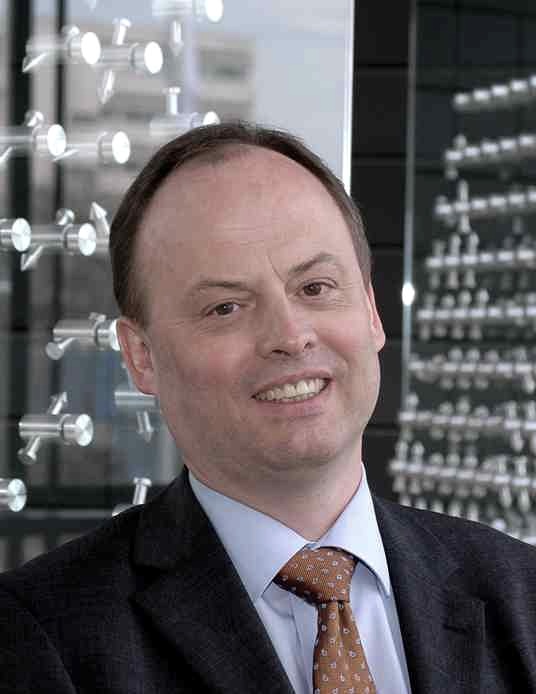}}]{Karlheinz Meier}
Karlheinz Meier received a Ph.D.\ in Physics from
Hamburg University, Germany, in 1984. He
is a Professor of Physics at Heidelberg University, Germany,
and Co-Founder of the Kirchhoff-Institut and the Heidelberg
ASIC Laboratory in Heidelberg. His research interests include the application
of microelectronics in particle physics. electronic realizations
of brain circuits and principles of information processing in spiking
neural networks.
\end{IEEEbiography}

\end{document}